\documentclass[twocolumn,dvipsnames]{aastex63}

\usepackage{makecell,CJK}
\usepackage{lineno}

\newcommand{\jwst}{{\it JWST}}

\providecommand\aap{Astron.\ Astrophys.} 
\providecommand\apjl{Astrophys.\ J.} 
\providecommand\apjs{Astrophys.\ J.\ Suppl.} 

\received{2024 September 19}
\accepted{2024 November 9}
\submitjournal{PSJ}

\shorttitle{Volatile Composition and Activity Evolution of 358P}
\shortauthors{Hsieh et al.}

\begin{document}
\begin{CJK*}{UTF8}{gbsn}

\title{The Volatile Composition and Activity Evolution of Main-Belt Comet 358P/PANSTARRS}

\correspondingauthor{Henry Hsieh}
\email{hhsieh@psi.edu}

\author[0000-0001-7225-9271]{Henry H.\ Hsieh}
\affiliation{Planetary Science Institute, 1700 East Fort Lowell Rd., Suite 106, Tucson, AZ 85719, USA}
\affiliation{Institute of Astronomy and Astrophysics, Academia Sinica, P.O.\ Box 23-141, Taipei 10617, Taiwan}

\author[0000-0003-2152-6987]{John W.\ Noonan}
\affiliation{Department of Physics, Auburn University, Edmund C.\ Leach Science Center, Auburn, 36849, AL, USA}

\author[0000-0002-6702-7676]{Michael S.\ P.\ Kelley}
\affiliation{Department of Astronomy, University of Maryland, 1113 Physical Sciences Complex, Building 415, College Park, MD 20742, USA}

\author[0000-0002-2668-7248]{Dennis Bodewits}
\affiliation{Department of Physics, Auburn University, Edmund C.\ Leach Science Center, Auburn, 36849, AL, USA}

\author[0000-0002-5736-1857]{Jana Pittichov\'a}
\affiliation{Jet Propulsion Laboratory, California Institute of Technology, 4800 Oak Grove Dr., Pasadena, CA 91109, USA}

\author[0000-0002-1506-4248]{Audrey Thirouin}
\affiliation{Lowell Observatory, 1400 W.\ Mars Hill Rd., Flagstaff, AZ 86001, USA}

\author[0000-0001-7895-8209]{Marco Micheli}
\affiliation{ESA PDO NEO Coordination Centre, Largo Galileo Galilei, 1, I-00044 Frascati (RM), Italy}

\author[0000-0003-2781-6897]{Matthew M.\ Knight}
\affiliation{Physics Department, U.S. Naval Academy, 572C Holloway Rd., Annapolis, MD, 21402, USA}
\affiliation{Department of Astronomy, University of Maryland, 1113 Physical Sciences Complex, Building 415, College Park, MD 20742, USA}

\author[0000-0003-3257-4490]{Michele T.\ Bannister}
\affiliation{School of Physical and Chemical Sciences - Te Kura Mat\=u, University of Canterbury, Christchurch 8041, New Zealand}

\author[0000-0001-7335-1715]{Colin O.\ Chandler}
\affiliation{Department of Astronomy \& the DiRAC Institute, University of Washington, 3910 15th Ave NE, Seattle, WA 98195, USA}
\affiliation{LSST Interdisciplinary Network for Collaboration and Computing, 933 N.\ Cherry Avenue, Tucson, AZ 85721, USA}
\affiliation{Department of Astronomy \& Planetary Science, Northern Arizona University, P.O.\ Box 6010, Flagstaff, AZ 86011, USA}
\affiliation{Raw Data Speaks Initiative, USA}

\author[0000-0002-4043-6445]{Carrie E.\ Holt}
\affiliation{Department of Astronomy, University of Maryland, 1113 Physical Sciences Complex, Building 415, College Park, MD 20742, USA}
\affiliation{Las Cumbres Observatory, 6740 Cortona Drive, Suite 102, Goleta, CA 93117, USA}

\author[0000-0001-6314-873X]{Matthew J.\ Hopkins}
\affiliation{School of Physical and Chemical Sciences - Te Kura Mat\=u, University of Canterbury, Christchurch 8041, New Zealand}

\author[0000-0002-9042-408X]{Yaeji Kim}
\affiliation{Department of Astronomy, University of Maryland, 1113 Physical Sciences Complex, Building 415, College Park, MD 20742, USA}

\author[0000-0001-6765-6336]{Nicholas A.\ Moskovitz}
\affiliation{Lowell Observatory, 1400 W.\ Mars Hill Rd., Flagstaff, AZ 86001, USA}

\author[0000-0001-5750-4953]{William J.\ Oldroyd}
\affiliation{Department of Astronomy \& Planetary Science, Northern Arizona University, P.O.\ Box 6010, Flagstaff, AZ 86011, USA}

\author[0009-0001-1692-4676]{Jack Patterson}
\affiliation{School of Physical and Chemical Sciences - Te Kura Mat\=u, University of Canterbury, Christchurch 8041, New Zealand}

\author[0000-0003-3145-8682]{Scott S.\ Sheppard}
\affiliation{Earth and Planets Laboratory, Carnegie Institution for Science, 5241 Broad Branch Road NW, Washington, DC 20015, USA}

\author[0000-0001-6541-8887]{Nicole Tan}
\affiliation{School of Physical and Chemical Sciences - Te Kura Mat\=u, University of Canterbury, Christchurch 8041, New Zealand}

\author[0000-0001-9859-0894]{Chadwick A.\ Trujillo}
\affiliation{Department of Astronomy \& Planetary Science, Northern Arizona University, P.O.\ Box 6010, Flagstaff, AZ 86011, USA}

\author[0000-0002-4838-7676]{Quanzhi Ye (叶泉志)}
\affiliation{Department of Astronomy, University of Maryland, 1113 Physical Sciences Complex, Building 415, College Park, MD 20742, USA}
\affiliation{Center for Space Physics, Boston University, 725 Commonwealth Ave, Boston, MA 02215, USA}

\begin{abstract}
We report the detection of water vapor associated with main-belt comet 358P/PANSTARRS on UT 2024 January 8-9 using the NIRSPEC instrument aboard \jwst{}. We derive a water production rate of $Q_{\rm H_2O}=(5.0\pm0.2)\times10^{25}$~molecules~s$^{-1}$, marking only the second direct detection of sublimation products of any kind from a main-belt comet, after 238P/Read.  
Similar to 238P, we find a remarkable absence of hypervolatile species, finding $Q_{\rm CO_2}<7.6\times10^{22}$~molecules~s$^{-1}$, corresponding to $Q_{\rm CO_2}/Q_{\rm H_{2}O}<0.2$\%.
Upper limits on CH$_{3}$OH and CO emission are also estimated.
Photometry from ground-based observations show that the dust coma brightened and faded slowly over $\sim250$~days in 2023-2024,
consistent with photometric behavior observed in 2012-2013, but also indicate a $\sim2.5\times$ decline in the dust production rate between these two periods. 
Dynamical dust modeling shows that the coma's morphology as imaged by \jwst{}'s NIRCAM instrument on 2023 November 22 can be reproduced by asymmetric dust emission from a nucleus with a mid-range obliquity ($\varepsilon\sim80^{\circ}$) with a steady-state mass loss rate of $\sim0.8$~kg~s$^{-1}$.
Finally, we find similar $Af\rho$-to-gas ratios of $\log_{10}(Af\rho/Q_{\rm H_2O})=-24.8\pm0.2$ for 358P and $\log_{10}(Af\rho/Q_{\rm H_2O})=-24.4\pm0.2$ for 238P, suggesting that $Af\rho$ could serve as an effective proxy for estimating water production rates in other active main-belt comets.  The confirmation of water vapor outgassing in both main-belt comets observed by \jwst{} to date reinforces the use of recurrent activity near perihelion as an indicator of sublimation-driven activity in active asteroids.
\end{abstract}




\section{Introduction\label{section:intro}}

\subsection{Background\label{section:background}}

Active asteroids are small Solar system objects that have asteroid-like orbits (as parameterized by the Tisserand parameter, or $T_J$) but exhibit visible mass loss \citep{jewitt2015_actvasts_ast4,jewitt2022_continuum_comets3}.  Meanwhile, main-belt comets (MBCs) are a subset of the active asteroids, encompassing objects whose mass loss is specifically driven by the sublimation of volatile ices \citep{hsieh2006_mbcs,snodgrass2017_mbcs}.

Until very recently, the sublimation-driven nature of MBC activity has been indirectly inferred from dust modeling demonstrating that the morphologies of active MBCs are consistent with sustained dust emission \citep[e.g.,][]{hsieh2009_238p,moreno2011_324p,licandro2013_288p,jewitt2014_133p}, and observations of recurring activity near perihelion with intervening periods of inactivity elsewhere in their orbits \citep[e.g.,][]{hsieh2004_133p,hsieh2009_238p,hsieh2018_238p288p}.  Both of these results are naturally explained by sublimation as the driver of activity and more difficult to explain as a consequence of other processes like rotational destabilization or impacts without the invocation of specifically contrived circumstances \citep{hsieh2012_scheila,jewitt2015_actvasts_ast4}.

For over 15 years after MBCs were first identified, numerous attempts were made to directly confirm the hypothesized sublimation-driven nature of MBC activity via spectroscopic searches for sublimation products associated with active MBCs \citep[see][]{snodgrass2017_mbcs}.  Many of these searches used the largest telescopes in the world --- the Very Large Telescope (VLT), the Gemini North and Gemini South observatories, the Gran Telescopio Canarias (GTC), and the Keck I Observatory --- to target emission from the CN $B~^2\Sigma^+ - X~^2\Sigma^+$ (0,0) band around 388.3~nm \citep[e.g.,][]{jewitt2009_259p,hsieh2012_288p,licandro2011_133p176p,snodgrass2017_mbcs}, which is commonly found and relatively easily detected in spectra of other comets \citep[e.g.,][]{rauer2003_halebopp,ahearn1995_ensemblecomets}.  None of these efforts resulted in the direct detection of CN, although they did place upper limits on CN production rates on the order of $Q_{\rm CN}\sim10^{23}$ -- $10^{24}$~molecules~s$^{-1}$, corresponding to water production rate limits on the order of $Q_{\rm H_2O}\sim10^{25}$ -- $10^{27}$~molecules~s$^{-1}$, assuming similar $Q_{\rm CN}/Q_{\rm H_2O}$ abundance ratios as have been found in previously measured comets \citep{ahearn1995_ensemblecomets}.  Attempts to directly detect water vapor or the photodissociation products of water from MBCs using the Heterodyne Instrument for the Far Infrared (HIFI) on the {\it Herschel Space Telescope} \citep[targeting the $1_{10}-1_{01}$ ground state rotational line of H$_2$O at 557 GHz;][]{devalborro2012_176pherschel,orourke2013_p2012t1} and XSHOOTER on the VLT \citep[targeting ultraviolet OH emission at 308~nm;][]{snodgrass2017_358p} were only able to set similar upper limits on water production rates.

Direct evidence of sublimation from a MBC was finally obtained when \jwst{} observations of 238P/Read (semimajor axis of $a=3.166$~au; eccentricity of $e=0.252$; inclination of $i=1.264^{\circ}$; $T_J=3.153$)\footnote{Orbit solution from the Jet Propulsion Laboratory (JPL) Small Body Database (SBDB; \url{https://ssd.jpl.nasa.gov/}) derived on 2024 May 30} on UT 2022 September 08 yielded a water vapor production rate of $Q_{\rm H_2O}=(9.9\pm1.0)\times10^{24}$~molecules~s$^{-1}$, or $Q_{\rm H_2O}=0.30\pm0.03$~kg~s$^{-1}$ \citep{kelley2023_jwst238p}, where 238P's orbital and observational geometric parameters at the time are listed in Table~\ref{table:jwst_mbc_observations}.
In this paper, we report the second-ever detection by \jwst{} of water vapor from a MBC, as well as results from additional analyses to characterize the contemporaneous production rates of dust and other hypervolatile species.

\setlength{\tabcolsep}{5pt}
\setlength{\extrarowheight}{0em}
\begin{table*}[htb!]
\caption{\jwst{} Observations of 238P and 358P}
\centering
\smallskip
\footnotesize
\begin{tabular}{cccrrrrrrrr}
\hline\hline
\multicolumn{1}{c}{Target}
 & \multicolumn{1}{c}{UT Date}
 & \multicolumn{1}{c}{Instrument}
 & \multicolumn{1}{c}{$\nu$$^a$}
 & \multicolumn{1}{c}{$r_h$$^b$}
 & \multicolumn{1}{c}{$\Delta$$^c$}
 & \multicolumn{1}{c}{$\alpha$$^d$}
 & \multicolumn{1}{c}{PA$_{-\odot}$$^e$}
 & \multicolumn{1}{c}{PA$_{-v}$$^f$}
 & \multicolumn{1}{c}{$\Delta t_q$$^g$}
 & \multicolumn{1}{c}{$\Delta t_{\rm NS}$$^h$}
 \\
\hline
238P & 2022-09-08 & NIRCam  & 28.3 & 2.428 & 2.091 & 24.6 & 261.1 & 259.5 & +95 & 0 \\
238P & 2022-09-08 & NIRSpec & 28.3 & 2.428 & 2.091 & 24.6 & 261.1 & 259.5 & +95 & --- \\
\hline
358P & 2023-11-22    & NIRCam  &  3.4 & 2.396 & 1.632 & 18.7 & 55.2 & 244.8 & +11 & --48 \\
358P & 2024-01-08 -- 2024-01-09 & NIRSpec & 17.4 & 2.416 & 2.183 & 24.3 & 67.9 & 242.8 & +59 & --- \\
\hline
\hline
\multicolumn{11}{l}{$^a$ True anomaly, in degrees} \\
\multicolumn{11}{l}{$^b$ Heliocentric distance, in au} \\
\multicolumn{11}{l}{$^c$ Geocentric distance, in au} \\
\multicolumn{11}{l}{$^d$ Solar phase angle (Sun-target-observer), in degrees} \\
\multicolumn{11}{l}{$^e$ Position angle of the anti-Solar vector as projected on the sky, in degrees East of North} \\
\multicolumn{11}{l}{$^f$ Position angle of the negative heliocentric velocity vector as projected on the sky,} \\
\multicolumn{11}{l}{$~~~$ in degrees East of North} \\
\multicolumn{11}{l}{$^g$ Time relative to perihelion (positive values indicating time after perihelion), in days.} \\
\multicolumn{11}{l}{$^h$ Time relative to NIRSpec observations (negative values indicating time prior to NIRSpec observations), in days.} \\
\end{tabular}
\label{table:jwst_mbc_observations}
\end{table*}

\subsection{358P/PANSTARRS\label{section:358p_background}}

358P/PANSTARRS ($a=3.147$~au; $e=0.239$; $i=11.060^{\circ}$; $T_J=3.136$)\footnote{Orbit solution from the JPL SSDB derived on 2024 March 5} was discovered as P/2012 T1 by the 1.8~m Pan-STARRS1 survey telescope on Haleakala \citep{wainscoat2012_p2012t1} on UT 2012 October 06, shortly after it passed perihelion on UT 2012 September 11. 
Optical imaging observations conducted between 2012 October and 2013 February showed photometric behavior that was judged to be qualitatively consistent with the action of a long-lasting dust emission driver such as sublimation \citep{hsieh2013_p2012t1}, an interpretation that was corroborated by an independent Monte Carlo dust modeling analysis \citep{moreno2013_p2012t1}.
Interestingly, the work by \citet{moreno2013_p2012t1} showed that 358P's morphology was best fit by anistropic dust emission from a mid-range obliquity ($\varepsilon\sim80^{\circ}$) object, with preferential dust ejection at high latitudes.
They specifically found that the object's brightness and morphology evolution was most consistent with a model of sustained anisotropic dust emission starting about two months prior to perihelion and ending four months after perihelion, i.e., from about July 2012 when the object was at a true anomaly of $\nu\sim345^{\circ}$, to January 2013 when the object was at $\nu\sim35^{\circ}$).  That analysis found a total ejected dust mass of $\sim(8-25)\times10^6$~kg, corresponding to an average dust production rate of ${\dot m}\sim(0.5-1.7)$~kg~s$^{-1}$ (depending on the choice of maximum particle sizes from $r_{\rm max}=1$~cm to $r_{\rm max}=10$~cm), peaking around 30-40 days after perihelion.  

The observational confirmation by \citet{hsieh2018_358p} that 358P exhibited recurrent activity near perihelion in 2017 strongly supported the conclusions of \citet{hsieh2013_p2012t1} and \citet{moreno2013_p2012t1} that the activity of this object is driven by sublimation.  In those 2017 observations, however, the object was seen to be active at $\nu\sim320^{\circ}$, or about three months earlier than the activity onset point for the object estimated by \citet{moreno2013_p2012t1} for its 2012-2013 active apparition.

The inactive nucleus of the object was determined by  \citet{hsieh2023_mbcnuclei} to have a $V$-band absolute magnitude of $H_V=20.2\pm0.3$~mag, corresponding to an effective radius of $r_N=0.3\pm0.1$~km, assuming a spherical nucleus with an $R$-band albedo of $p_V=0.05$.  Meanwhile, 10~hours of time-series observations with the VLT were inconclusive in terms of constraining a rotation period, though they marginally suggested a possible rotation period of $\sim8$~hr \citep{agarwal2018_358pnucleus}.

Multiple searches for sublimation products using the methods described in Section~\ref{section:background} specifically targeting 358P have been made, all resulting in non-detections.
\citet{hsieh2013_p2012t1}
found an upper limit CN production rate of $Q_{\rm CN}=1.5\times10^{23}$~molecules~s$^{-1}$ (which was inferred to correspond to a H$_2$O production rate of $Q_{\rm H_2O}<5\times10^{25}$~molecules~s$^{-1}$) from Keck Low-Resolution Imaging Spectrometer observations, while
\citet{orourke2013_p2012t1} found an upper limit H$_2$O production rate of $Q_{\rm H_2O}=7.63\times10^{25}$~molecules~s$^{-1}$ from {\it Herschel} observations, and
\citet{snodgrass2017_358p} found an upper limit OH production rate of $Q_{\rm OH}=6\times10^{25}$~molecules~s$^{-1}$ (corresponding to an inferred H$_2$O production rate of $Q_{\rm H_2O}<9\times10^{25}$~molecules~s$^{-1}$) from VLT XSHOOTER observations.

Dynamical analyses by \citet{hsieh2013_p2012t1} indicated that 358P was unlikely to be a recently implanted interloper from the outer Solar system, and in fact, is dynamically linked to the $\sim155$~Myr old Lixiaohua asteroid family.  Notably, the Lixiaohua family is also associated with another MBC, 313P/Gibbs \citep{hsieh2015_313p,hsieh2018_activeastfamilies}, where a spectroscopic study of a large sample of Lixiaohua family members led \citet{depra2020_lixiaohua} to conclude that the available compositional evidence supports the classification of both  313P and 358P as ``true'' members of this family.

\section{Observations\label{section:observations}}

\jwst{} \citep{gardner2023_jwst} observations of 358P were obtained by NIRCam
\citep{rieke2023_nircam}
on UT 2023 November 22 and by NIRSpec 
\citep{jakobsen2022_nirspec,boker2023_nirspec}
on UT 2024 January 8--9 (Table~\ref{table:jwst_mbc_observations}) as part of \jwst{} General Observer (GO) program GO 4250\footnote{\url{https://www.stsci.edu/jwst/science-execution/program-information?id=4250}} in Cycle 2.  Observational circumstances of these observations are listed in Table~\ref{table:jwst_mbc_observations}.

\setlength{\tabcolsep}{8pt}
\setlength{\extrarowheight}{0em}
\begin{table*}[htb!]
\caption{Ground-Based Observing Instrumentation Characteristics}
\centering
\smallskip
\footnotesize
\begin{tabular}{lcccc}
\hline\hline
\multicolumn{1}{c}{Telescope$^a$}
 & \multicolumn{1}{c}{Instrument}
 & \multicolumn{1}{c}{FOV$^b$}
 & \multicolumn{1}{c}{Pixel Scale}
 & \multicolumn{1}{c}{Binning}
 \\[2pt]
\hline
ARC & ARCTIC & $7\farcm5\times7\farcm5$ & $0\farcs228$ & $1\times1$ \\
Magellan & IMACS   & $15\farcm4\times15\farcm4$ & $0\farcs20$ & $1\times1$ \\
Gemini-N & GMOS-N  & $5\farcm5\times5\farcm5$ & $0\farcs16$ & $2\times2$ \\
Gemini-S & GMOS-S  & $5\farcm5\times5\farcm5$ & $0\farcs16$ & $2\times2$ \\
Palomar & WaSP    & $18\farcm4\times18\farcm5$ & $0\farcs175$ & $1\times1$ \\
LDT (2x)  & LMI & $12\farcm3\times12\farcm3$ & $0\farcs24$ & $2\times2$ \\
LDT (3x) & LMI & $12\farcm3\times12\farcm3$ & $0\farcs36$ & $3\times3$ \\
Mt.\ John & FLI CCD & $24\farcm6\times24\farcm6$ & $0\farcs72$ & $1\times1$ \\
SOAR & Goodman & $7\farcm2\times7\farcm2$ & $0\farcs30$ & $2\times2$ \\
\hline
\hline
\multicolumn{5}{l}{$^a$ ARC: Astrophysical Research Consortium 3.5~m telescope;} \\
\multicolumn{5}{l}{$~~~$ Magellan: Magellan Baade telescope; Gemini-N: Gemini North} \\
\multicolumn{5}{l}{$~~~$ telescope; Gemini-S: Gemini South telescope; Palomar: Palomar } \\
\multicolumn{5}{l}{$~~~$ Hale Telescope; LDT (2x): Lowell Discovery Telescope in 2$\times$2} \\
\multicolumn{5}{l}{$~~~$ binning mode; LDT (3x): Lowell Discovery Telescope in 3$\times$3} \\
\multicolumn{5}{l}{$~~~$ binning mode; Mt.\ John: Mt.\ John Boller \& Chivens Telescope; } \\
\multicolumn{5}{l}{$~~~$ SOAR: Southern Astrophysical Research Telescope } \\
\multicolumn{5}{l}{$^b$ Field of view} \\
\end{tabular}
\label{table:instrumentation}
\end{table*}

NIRCam observations of 358P using the F200W and F277W broadband filters were obtained simultaneously using two separate detectors and a dichroic. Both detectors have fields of view that are $2040\times2048$ pixels in size.  The pixel scales are $0\farcs0313$~pixel$^{-1}$ and $0\farcs0630$~pixel$^{-1}$ for the F200W and F277W images, respectively, providing angular fields of view of $63\farcs9\times64\farcs1$ for F200W images and $128\farcs5\times129\farcs0$ for F277W images.  For a Solar spectrum \citep{wilmer2018_solarspectrum}, the F200W and F277W filters (which nominally cover wavelengths from 1.725~$\mu$m to 2.260~$\mu$m, and from 2.367~$\mu$m to 3.220~$\mu$m, respectively) have effective wavelengths of 1.97~$\mu$m and 2.74~$\mu$m, respectively \citep{kelley2023_jwst238p}.  Four exposures were taken using the 4-point INTRAMODULEBOX dither pattern \citep[see][]{coe2017_nircamdithers}, which enables mitigation of detector artifacts and cosmic rays by applying small pointing offsets between exposures.  Each detector was read out using the SHALLOW4 readout pattern for a total exposure time of 1031~s per filter.

Meanwhile, NIRSpec observations of 358P and an off-source background field were obtained using the instrument's integral field unit (IFU), which provides spatially resolved imaging spectroscopy over a $3''\times3''$ field of view producing an IFU data cube with spatial elements that are $0\farcs1\times0\farcs1$ in size.  Observations utilized the IFU's PRISM/CLEAR mode which delivers a nominal resolving power of $30-330$ over a wavelength range of 0.6~$\mu$m -- 5.3~$\mu$m \citep{boker2023_nirspec}. Four exposures were taken using the 4-POINT-DITHER dither pattern and read out using the NRSIRS2RAPID readout pattern for a total exposure time of 2976~s.

Finally, supporting ground-based observations were obtained with
the Astrophysical Research Consortium (ARC) Telescope Imaging Camera (ARCTIC) on the ARC 3.5~m telescope at Apache Point Observatory in New Mexico, USA;
the Inamori Magellan Areal Camera and Spectrograph \citep[IMACS;][]{dressler2011_imacs} on the 6.5~m Magellan-Baade telescope at Las Campanas in Chile;
the Gemini Multi-Object Spectrograph - North \citep[GMOS-N;][]{hook2004_gmos} in imaging mode on the 8.1~m Gemini North (Gemini-N) telescope (program GN-2022B-Q-307) on Maunakea in Hawaii, USA;
the Gemini Multi-Object Spectrograph - South \citep[GMOS-S;][]{gimeno2016_gmoss} in imaging mode on the 8.1~m Gemini South (Gemini-S) telescope (programs GS-2022B-LP-104 and GS-2022B-Q-111) at Cerro Pach{\'o}n in Chile;
the Large Monolithic Imager \citep[LMI;][]{bida2014_dct} on Lowell Observatory's 4.3~m Lowell Discovery Telescope (LDT; formerly named the Discovery Channel Telescope) at Happy Jack, Arizona, USA;
the Wafer-Scale camera for Prime \citep[WaSP;][]{nikzad2017_detectors} wide field prime focus camera
on the 5.1~m Hale Telescope at Palomar Observatory in California, USA;
a FLI CCD camera on the 0.61~m Boller \& Chivens Telescope at the University of Canterbury Mount John Observatory near Lake Tekapo, New Zealand; and
the Goodman High Throughput Spectrograph Blue Camera \citep{clemens2004_goodman} on the 4.1~m Southern Astrophysical Research Telescope (SOAR; program 2023A-396684) at Cerro Pach{\'o}n in Chile.
All ground-based observations reported here were obtained using Sloan $r'$-band filters.
Details of all instrumentation are shown in Table~\ref{table:instrumentation}, while observational circumstances of all ground-based observations are listed in Table~\ref{table:ground_observations_358p}.

\setlength{\tabcolsep}{4.5pt}
\setlength{\extrarowheight}{0em}
\begin{table*}[htb!]
\caption{Ground-based $r'$-band Observations of 358P}
\centering
\smallskip
\footnotesize
\begin{tabular}{ccrrrrrrrrrrr}
\hline\hline
\multicolumn{1}{c}{UT Date}
 & \multicolumn{1}{c}{Telescope$^a$}
 & \multicolumn{1}{c}{$N$$^b$}
 & \multicolumn{1}{c}{$t$$^c$}
 & \multicolumn{1}{c}{$\theta_s$$^d$}
 & \multicolumn{1}{c}{$\nu$$^f$}
 & \multicolumn{1}{c}{$r_h$$^g$}
 & \multicolumn{1}{c}{$\Delta$$^h$}
 & \multicolumn{1}{c}{$\alpha$$^i$}
 & \multicolumn{1}{c}{PA$_{-\odot}$$^j$}
 & \multicolumn{1}{c}{PA$_{-v}$$^k$}
 & \multicolumn{1}{c}{$\Delta t_q$$^l$}
 & \multicolumn{1}{c}{$\Delta t_{\rm NS}$$^m$}
 \\
\hline
2023 Jun 30 & SOAR$^*$ & 5 & 600 & 0.7 & 321.3 & 2.501 & 2.353 & 23.9 & 246.1 & 243.7 & --134 & --193 \\
2023 Jul 26 & ARC   & 4 & 600 & 1.4 & 328.8 & 2.464 & 2.007 & 23.5 & 251.2 & 242.4 & --108 & --167 \\
2023 Sep 12 & Palomar   & 14 & 4200 & 1.3 & 342.5 & 2.416 & 1.529 & 14.2 & 273.3 & 242.1 & --60 & --119 \\
2023 Oct 10 & Palomar   & 10 & 3000 & 1.4 & 350.7 & 2.401 & 1.430 &  7.2 & 337.2 & 243.6 & --32 & --91 \\
2023 Oct 16 & LDT (2x)  & 3 &  900 & 1.2 & 352.4 & 2.399 & 1.433 & 7.6 & 359.9 & 243.9 & --26 & --85 \\
2023 Oct 19 & LDT (2x)  & 3 & 600 & 1.1 & 353.3 & 2.398 & 1.438 &  8.2 &  9.7  & 244.1 & --23 & --82 \\
2023 Nov 04 & Magellan  & 1 &  300 & 1.0 & 358.0 & 2.395 & 1.500 & 12.9 & 40.7 & 244.7 & --7 & --66 \\
2023 Nov 06 & LDT (3x)  & 8 & 1560 & 1.3 & 358.6 & 2.395 & 1.512 & 13.6 & 43.1 & 244.8 & --5 & --64\\
2023 Nov 11 & {\it Perihelion} & -- & -- & -- & 0.0 & 2.395 & 1.543 & 12.8 & 47.5 & 244.9 & 0 & --59 \\
2023 Nov 21 & LDT (3x)  & 2 & 600 & 3.1 &   3.1 & 2.396 & 1.626 & 18.1 & 54.4 & 244.8 & +10 & --49 \\
2023 Dec 08 & LDT (2x)  & 9 & 3100 & 1.7 &   8.1 & 2.399 & 1.796 & 21.6 &  61.2 & 244.3 & +27 & --32 \\
2023 Dec 21 & LDT (2x)  & 33 & 3300 & 1.3 &  11.9 & 2.405 & 1.948 & 23.2 &  64.4 & 243.6 & +40 & --19 \\
2024 Jan 12 & Mt.\ John & 7 & 2100 & 4.2 & 18.4 & 2.419 & 2.231 & 24.0 & 68.2 & 242.7 & +62 & +3 \\
2024 Feb 01 & LDT (3x)  & 4 & 1200 & 1.2 &  24.1 & 2.436 & 2.489 & 23.0 & 70.8 & 242.4 & +82 & +23 \\
2024 Feb 02 & Gemini-N  & 3 &  900 & 0.6 &  24.5 & 2.437 & 2.503 & 23.0 &  70.9 & 242.4 & +83 & +24 \\
2024 Feb 27 & Gemini-N  & 3 &  900 & 1.1 &  31.5 & 2.465 & 2.818 & 20.2 & 74.0 & 243.0 & +108 & +49 \\
2024 Mar 02 & LDT (2x)  & 2  & 600 & 1.8 &  32.6 & 2.470 & 2.864 & 19.7 & 74.5 & 243.2 & +112 & +53 \\
\hline
\hline
\multicolumn{13}{l}{$^a$ See Table~\ref{table:instrumentation} for telescope name explanations} \\
\multicolumn{13}{l}{$^b$ Number of usable exposures.} \\
\multicolumn{13}{l}{$^c$ Total exposure time, in s.} \\
\multicolumn{13}{l}{$^d$ FWHM seeing, in arcseconds.} \\
\multicolumn{13}{l}{$^e$ Angular extent corresponding to 5000~km at the distance of the target, in arcseconds.} \\
\multicolumn{13}{l}{$^f$ True anomaly, in degrees.} \\
\multicolumn{13}{l}{$^g$ Heliocentric distance, in au.} \\
\multicolumn{13}{l}{$^h$ Geocentric distance, in au.} \\
\multicolumn{13}{l}{$^i$ Solar phase angle (observer-target-Sun), in degrees.} \\
\multicolumn{13}{l}{$^j$ Position angle of the anti-Solar vector as projected on the sky, in degrees East of North.} \\
\multicolumn{13}{l}{$^k$ Position angle of the negative heliocentric velocity vector as projected on the sky, in degrees East of North.} \\
\multicolumn{13}{l}{$^l$ Time prior to (negative values) or after (positive values) perihelion, in days.} \\
\multicolumn{13}{l}{$^m$ Time prior to (negative values) or after (positive values) NIRSpec observations, in days.} \\
\multicolumn{13}{l}{$^*$ Target not detected at expected position} \\
\end{tabular}
\label{table:ground_observations_358p}
\end{table*}

For later comparison of our 358P results to those reported by \citet{kelley2023_jwst238p} for 238P, we also report previously unpublished supporting ground-based observations of 238P using Gemini North and GMOS-N, Gemini South and GMOS-S, and the Hale Telescope and WaSP obtained between UT 2022 August 31 and UT 2022 October 01, bracketing \jwst{}'s NIRCam and NIRSpec observations on UT 2022 September 08 and, in particular, including Gemini North observations obtained on the same date as \jwst{}'s observations.  Observational circumstances of these ground-based observations are listed in Table~\ref{table:ground_observations_238p}.

\setlength{\tabcolsep}{4.5pt}
\setlength{\extrarowheight}{0em}
\begin{table*}[htb!]
\caption{Ground-based $r'$-band Observations of 238P$^a$}
\centering
\smallskip
\footnotesize
\begin{tabular}{ccrrcrrrrrrrr}
\hline\hline
\multicolumn{1}{c}{UT Date}
 & \multicolumn{1}{c}{Telescope}
 & \multicolumn{1}{c}{$N$}
 & \multicolumn{1}{c}{$t$}
 & \multicolumn{1}{c}{$\theta_s$}
 & \multicolumn{1}{c}{$\nu$}
 & \multicolumn{1}{c}{$r_h$}
 & \multicolumn{1}{c}{$\Delta$}
 & \multicolumn{1}{c}{$\alpha$}
 & \multicolumn{1}{c}{PA$_{-\odot}$}
 & \multicolumn{1}{c}{PA$_{-v}$}
 & \multicolumn{1}{c}{$\Delta t_q$$^l$}
 & \multicolumn{1}{c}{$\Delta t_{\rm NS}$$^m$}
 \\
\hline
2022 Jun 05 & {\it Perihelion} & -- & -- & -- & 0.0 & 2.369 & 3.023 & 25.1 & 249.3 & 248.7 & 0 & --95 \\ 
2022 Aug 31 & Gemini-N &  3 &  450 & 0.5 & 25.8 & 2.418 & 2.180 & 24.7 & 260.1 & 258.6 & +87 & --8 \\
2022 Sep 03 & Gemini-N & 18 & 2700 & 0.5 & 26.7 & 2.421 & 2.147 & 24.6 & 260.5 & 258.9 & +90 & --5 \\
2022 Sep 08 & Gemini-N & 18 & 2700 & 0.6 & 28.1 & 2.427 & 2.093 & 24.4 & 260.9 & 259.4 & +95 & 0 \\
2022 Sep 22 & Gemini-N &  7 & 1050 & 0.6 & 32.1 & 2.445 & 1.945 & 23.0 & 262.0 & 260.5 & +109 & +14 \\
2022 Sep 22 & Gemini-S &  1 &  150 & 0.9 & 32.1 & 2.445 & 1.945 & 23.0 & 262.0 & 260.5 & +109 & +14 \\
2022 Sep 30 & Palomar  & 3 &  900 & 1.1 & 34.4 & 2.456 & 1.865 & 21.8 & 262.4 & 260.9 & +117 & +22 \\
2022 Oct 01 & Gemini-N & 2 &  300 & 0.7 & 34.7 & 2.457 & 1.855 & 21.6 & 262.4 & 261.0 & +118 & +23 \\
\hline
\hline
\multicolumn{13}{l}{$^a$ See Table~\ref{table:ground_observations_238p} for explanations of column headings} \\
\end{tabular}
\label{table:ground_observations_238p}
\end{table*}

All ground-based observations were conducted using non-sidereal tracking and at airmasses of $\lesssim2.5$, with typical seeing conditions of $\sim1''-2''$.  A minimum of three exposures was obtained during each visit in order to ensure that our target and any associated activity could be unambiguously identified from their non-sidereal motion.  In some cases, however, certain detections in a sequence were discarded due to being too close to background sources for photometry to be reliable (e.g., in Magellan observations on UT 2023 November 04), leading to fewer than 3 exposures per target on a night being reported here.

\section{Data Reduction\label{section:data_reduction}}

\subsection{\jwst{} NIRSpec Data Reduction\label{section:nirspec_data_reduction}}

During NIRSpec observations of 358P, observations of off-source background sky 1\arcmin{} away were also obtained for subtraction using the \jwst{} {\tt python} package and the {\tt spec2} data reduction pipeline\footnote{\url{https://jwst-pipeline.readthedocs.io/}} \citep{bushouse2023_jwstcalibrationpipeline_1_12_5}. During the fourth of four dithers for the target observations, the fine guidance sensor was found to have tracked on a hot pixel rather than the guide star, smearing the comet in the NIRSpec frame. The data in that dither was unrecoverable, and therefore omitted from further processing and analysis. During this process, the background spectrum was subtracted from the IFU data; the extended coma of the comet did not guarantee that a dust and/or gas free background spectrum could be obtained within the $3\arcsec\times3\arcsec$ FOV of the on-source data. The {\tt spec2} pipeline was run using version 1.12.5, allowing the NSClean step to be executed to reduce the $1/f$ noise feature \citep{rauscher2024_nsclean}.
At this stage, the data were photometrically calibrated to radiance units of MJy~sr$^{-1}$.

\begin{figure}[ht]
    \centering
    \includegraphics[width=\linewidth]{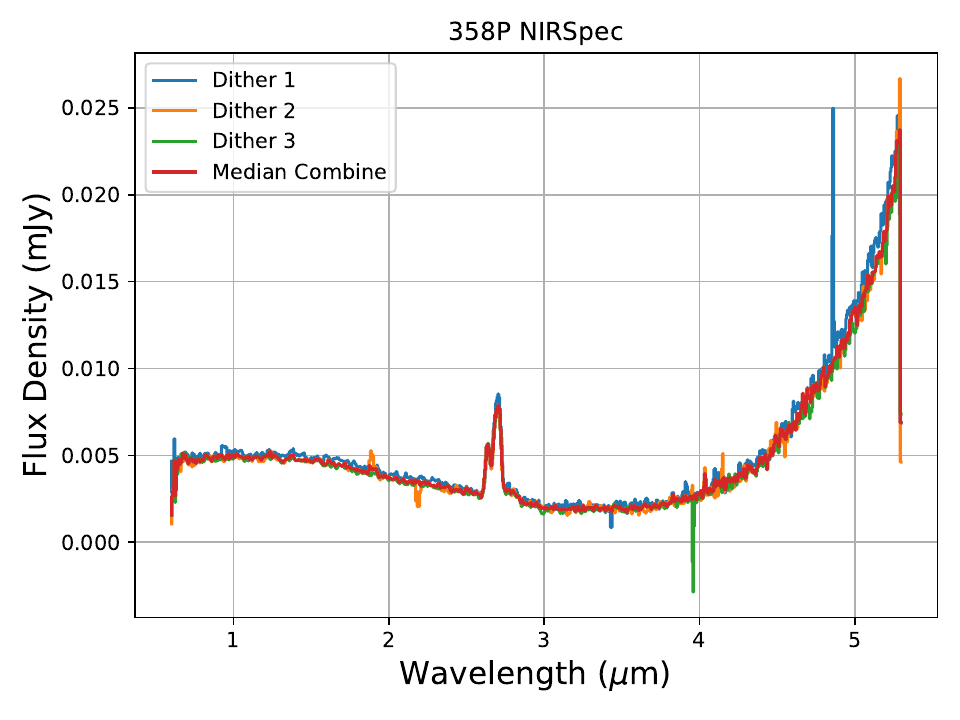}
    \caption{One-dimensional background-subtracted NIRSpec spectra for each dither extracted using $0\farcs4$-radius apertures, along with the median-combined spectrum.}
    \label{fig:NIRSpec_extracted}
\end{figure}

A one-dimensional spectrum was extracted from each dither using an aperture with a $0\farcs4$ radius (Figure~\ref{fig:NIRSpec_extracted}). This was accomplished using a slice by slice method in {\tt python} centered on the brightest pixel near the comet's predicted location in the frame. This method mitigated the effect of any small pixel-scale changes to the optocenter that occurred as a function of wavelength for the IFU. To avoid contamination of the optocenter determination by extended ``snowball'' cosmic ray events \citep{regan2024_snowballs}, which were occasionally missed by prior cosmic ray cleaning algorithms, a separate algorithm was used to find all pixels $50\sigma$ above the median signal of their immediate neighbors. These pixels then had their values replaced with the median of their nearest neighbors. This algorithm had the benefit of identifying the extended cosmic ray events without removing excess dust coma signal. Following these two custom steps, we found that the one-dimensional spectra extracted from the dataset were nearly identical in absolute flux and ideal for stacking to search for weak gas emissions.

\begin{figure*}[ht]
    \centering
    \includegraphics[width=0.8\linewidth]{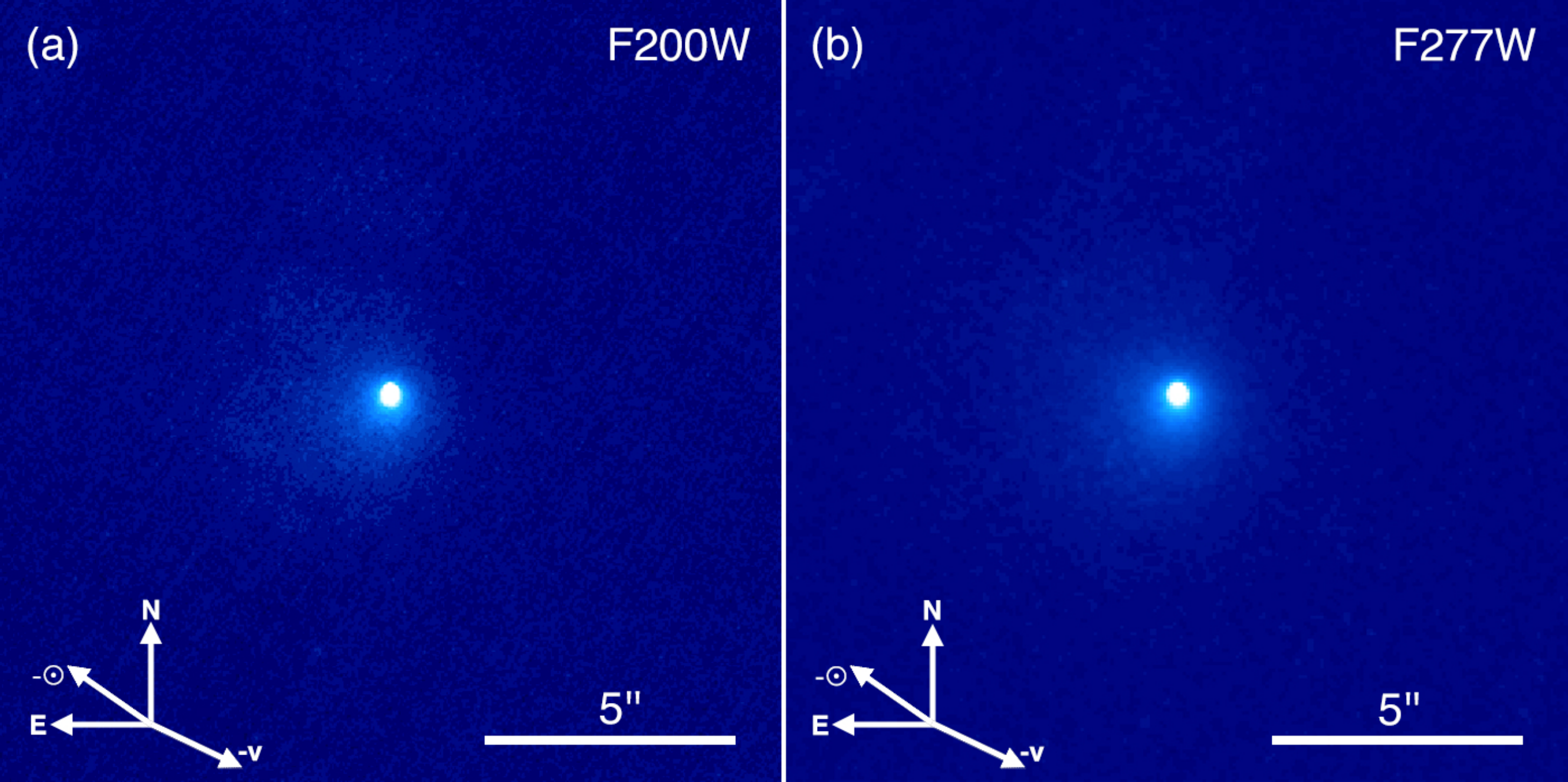}
    \caption{Median composite images of 358P/PANSTARRS, aligned on the photocenter of the comet in each individual image, constructed from NIRCam data obtained using the (a) F200W and (b) F277W broadband filters, comprising 1031~s of total exposure time each.  Labeled arrows indicate the directions of celestial north (N) and east (E), and the projected anti-Sun ($-\odot$) and negative heliocentric velocity ($-v$) vectors as seen from \jwst{}. A $5''$ angular scale bar (7600 km at the distance of the comet) is also shown in each panel. Color scaling in both panels is linear, where regions in the inner coma (in the center of each image) that are shown as solid white consist of pixels with fluxes that are $\sim20$\% of the peak central flux or larger.}
    \label{fig:nircam_images}
\end{figure*}

\subsection{\jwst{} NIRCam Data Reduction\label{section:nircam_data_reduction}}

NIRCam images were processed with pipeline version 1.11.4 and Calibration Reference Data System context file number 1149.  
At the time of these observations, individual pipeline-processed NIRCam images had a visible horizontal ``striped'' background structure that was attributed to $1/f$ noise from \jwst{}'s SIDECAR ASICs detector readout electronics\footnote{\url{https://jwst-docs.stsci.edu/known-issues-with-jwst-data/nircam-known-issues/nircam-1-f-noise-removal-methods}}, as well as relatively large numbers of randomly distributed pixels with {\tt NaN} values.

To reduce the impact of these noise features on subsequent analyses, we first addressed the $1/f$ noise issue by using the {\tt image1overf} package developed by C.\ Willott\footnote{\url{https://github.com/chriswillott/jwst}}, which removed row median levels while preserving overall flux levels.
We then processed the resulting images using our own {\tt python} code to replace {\tt NaN} pixel values (which comprised $\sim1$\% of all pixels in the uncorrected images) and uncertainties with the median value and uncertainty of all adjacent non-{\tt NaN} pixels. In cases of clusters of {\tt NaN} pixels, this process could result in new pixel values being computed from adjacent pixels whose values had to be computed themselves from other pixels, leading to increasingly meaningless pixel values for larger {\tt NaN} pixel clusters.  However, no particularly large {\tt NaN} pixel clusters occurred near our target in any of our NIRCam images, and so the impact of this effect on our subsequent analyses was minimal.

To maximize signal-to-noise ratios to characterize the detailed morphology of the comet and to mitigate the effect of cosmic rays and detector artifacts, we used {\tt pyraf}\footnote{\url{https://pypi.org/project/pyraf/}} \citep{stsci2012_pyraf} to construct median composite images of the object in each filter. This was done by shifting and aligning individual images on the object's photocenter using linear interpolation and performing a median combination of the resulting images for each filter.  Final median composite images are shown in Figure~\ref{fig:nircam_images}.

\subsection{Optical Data Reduction\label{section:optical_data_reduction}}

Standard bias subtraction, flatfield correction, and cosmic ray removal were performed for all optical images obtained from ground-based facilities using {\tt python} code utilizing the {\tt ccdproc} package\footnote{\url{https://ccdproc.readthedocs.io/}} \citep{craig2023_ccdproc} in {\tt astropy}\footnote{\url{http://www.astropy.org}} \citep{astropy2018_astropy} and the {\tt L.A.Cosmic} {\tt python} module\footnote{\url{https://pypi.org/project/lacosmic/}} \citep{vandokkum2001_lacosmic,vandokkum2012_lacosmic}.
Photometry measurements of the target object and at least one background reference star were performed using {\tt IRAF} \citep{tody1986_iraf,tody1993_iraf,fitzpatrick2024_iraf} and {\tt pyraf} software,
where photometry of reference stars was obtained by measuring net fluxes within circular apertures with sizes chosen using curve-of-growth analyses of representative stars, with background sampled from surrounding circular annuli.
Photometry of target objects was performed using circular apertures with fixed angular radii of $\rho=4''$, as well as circular apertures with radii of $\rho=5000$~km at the distance of the comet, where background statistics were measured in nearby but non-adjacent regions of blank sky to avoid potential dust contamination from the object or nearby field stars.

Absolute photometric calibration was performed using field star magnitudes from the ATLAS Refcat2 all-sky stellar reference catalog \citep{tonry2018_refcat}, where conversion of non-SDSS photometry to magnitudes in the SDSS system was accomplished as needed using transformations derived by \citet{tonry2012_ps1} and by R.~Lupton\footnote{{\url{http://www.sdss.org/}}}. 
We aimed to use $5-30$ well-isolated reference stars (i.e., field stars with no other neighboring sources within the photometry aperture used for those data, and ideally, within the annuli used to measure sky background as well) for photometric calibration where possible.
In some cases, however, only a few suitable reference stars, or even just one, were available due to the small margin between the limiting magnitude of the Refcat2 catalog and the saturation limit of many of our observations.

\setlength{\tabcolsep}{5pt}
\setlength{\extrarowheight}{0em}
\begin{table*}[htb!]
\caption{$r'$-band Photometric Results for 358P}
\centering
\smallskip
\footnotesize
\begin{tabular}{ccccccccc}
\hline\hline
\multicolumn{1}{c}{UT Date}
 & \multicolumn{1}{c}{Telescope$^a$}
 & \multicolumn{1}{c}{$m_{r,4''}$$^b$}
 & \multicolumn{1}{c}{$m_{V,4''}(1,1,0)$$^c$}
 & \multicolumn{1}{c}{$Af\rho_{4''}$$^d$}
 & \multicolumn{1}{c}{$\rho_{5000{\rm km}}$$^e$}
 & \multicolumn{1}{c}{$m_{r,5000{\rm km}}$$^f$}
 & \multicolumn{1}{c}{$m_{V,5000{\rm km}}(1,1,0)$$^g$}
 & \multicolumn{1}{c}{$Af\rho_{5000{\rm km}}$$^h$}
 \\
\hline
2023 Jun 30$^*$ & SOAR  & $>23.8$ & $>19.0$ & $<0.7$ & --- & --- & --- & --- \\
2023 Jul 26 & ARC   & 23.2$\pm$0.1   & 19.0$\pm$0.1   & 0.7$\pm$0.3 & 3.4 & 23.1$\pm$0.1 & 18.9$\pm$0.1 & 0.7$\pm$0.3 \\
2023 Sep 12 & Palomar   & 20.92$\pm$0.05 & 17.68$\pm$0.06 & 3.4$\pm$1.1 & 4.5 & 20.87$\pm$0.07 & 17.64$\pm$0.08 & 3.2$\pm$0.9 \\
2023 Oct 10 & Palomar   & 19.92$\pm$0.01 & 17.10$\pm$0.02 & 6.1$\pm$1.8 & 4.8 & 19.85$\pm$0.01 & 17.03$\pm$0.02 & 5.5$\pm$1.6  \\
2023 Oct 16 & LDT (2x)  & 19.75$\pm$0.01 & 16.92$\pm$0.01 & 7.1$\pm$2.1 & 4.8 & 19.69$\pm$0.01 & 16.85$\pm$0.02 & 6.5$\pm$1.9 \\
2023 Oct 19 & LDT (2x)  & 19.75$\pm$0.03 & 16.88$\pm$0.03 & 7.5$\pm$2.2 & 4.8 & 19.69$\pm$0.03 & 16.82$\pm$0.03 & 6.8$\pm$2.0 \\
2023 Nov 04 & Magellan  & 19.73$\pm$0.01 & 16.61$\pm$0.01 & 9.3$\pm$2.7 & 4.6 & 19.67$\pm$0.01 & 16.54$\pm$0.01 & 8.6$\pm$2.5 \\
2023 Nov 06 & LDT (3x)  & 19.76$\pm$0.02 & 16.60$\pm$0.02 & 9.4$\pm$2.7 & 4.5 & 19.69$\pm$0.02 & 16.52$\pm$0.02 & 8.8$\pm$2.5 \\
2023 Nov 21 & LDT (3x)  & 20.08$\pm$0.01 & 16.62$\pm$0.01 & 8.7$\pm$2.5 & 4.2 & 20.02$\pm$0.01 & 16.56$\pm$0.01 & 8.4$\pm$2.5 \\
2023 Dec 08 & LDT (2x)  & 20.17$\pm$0.01 & 16.39$\pm$0.02 & 9.4$\pm$1.6 & 3.8 & 20.20$\pm$0.01 & 16.42$\pm$0.02 & 9.6$\pm$1.6 \\
2023 Dec 21 & LDT (2x)  & 20.31$\pm$0.01 & 16.31$\pm$0.01 & 9.5$\pm$1.6 & 3.5 & 20.39$\pm$0.01 & 16.39$\pm$0.01 & 9.9$\pm$1.6 \\
2024 Jan 12 & Mt.\ John & 21.2$\pm$0.1 & 16.7$\pm$0.1 & 7.1$\pm$1.2 & 3.1 & 21.2$\pm$0.1 & 16.9$\pm$0.1 & 7.6$\pm$1.3 \\
2024 Feb 01 & LDT (3x)  & 20.87$\pm$0.03 & 16.32$\pm$0.03 & 7.5$\pm$1.2 & 2.8 & 21.15$\pm$0.04 & 16.63$\pm$0.04 & 8.3$\pm$1.4 \\
2024 Feb 02 & Gemini-N  & 20.98$\pm$0.01 & 16.41$\pm$0.01 & 6.9$\pm$1.1 & 2.7 & 21.23$\pm$0.01 & 16.66$\pm$0.02 & 7.7$\pm$1.3 \\
2024 Feb 27 & Gemini-N  & 21.16$\pm$0.11 & 16.39$\pm$0.11 & 6.1$\pm$1.1 & 2.4 & 21.56$\pm$0.06 & 16.79$\pm$0.06 & 7.0$\pm$1.2 \\
2024 Mar 02 & LDT (2x)  & 21.15$\pm$0.04 & 16.35$\pm$0.04 & 6.3$\pm$1.0 & 2.4 & 21.65$\pm$0.04 & 16.85$\pm$0.04 & 7.3$\pm$1.2 \\
\hline
\hline
\multicolumn{9}{l}{$^a$ See Table~\ref{table:ground_observations_358p} for explanations of telescope designations.} \\
\multicolumn{9}{l}{$^b$ Mean $r'$-band magnitude using $\rho=4\farcs0$ photometry apertures.} \\
\multicolumn{9}{l}{$^c$ Mean absolute $V$-band magnitude (normalized to $r_h=\Delta=1$~au and $\alpha=0^{\circ}$) using $\rho=4\farcs0$ photometry apertures.} \\
\multicolumn{9}{l}{$^d$ $Af\rho$, in cm, for $\rho=4\farcs0$ computed from best-fit power law to measured $Af\rho$ vs.\ aperture radius function, with uncertainties} \\
\multicolumn{9}{l}{$~~~~$indicating the range of $Af\rho$ values for $\rho=4\farcs0\pm1\farcs6$.} \\
\multicolumn{9}{l}{$^e$ Angular equivalent to a $\rho=5000$~km photometry aperture at the geocentric distance of the comet} \\
\multicolumn{9}{l}{$^f$ Mean $r'$-band magnitude using $\rho=5000$~km photometry apertures.} \\
\multicolumn{9}{l}{$^g$ Mean absolute $V$-band magnitude (normalized to $r_h=\Delta=1$~au and $\alpha=0^{\circ}$) using $\rho=5000$~km photometry apertures.} \\
\multicolumn{9}{l}{$^h$ $Af\rho$, in cm, for $\rho=5000$~km computed from best-fit power law to measured $Af\rho$ vs.\ aperture radius function, with} \\
\multicolumn{9}{l}{$~~~~$uncertainties indicating the range of $Af\rho$ values for $\rho=(5000\pm2000)$~km.} \\
\multicolumn{9}{l}{$^*$ $3\sigma$ point source detection limits for $\rho=1\farcs8$ photometry apertures used for field stars.}
\end{tabular}
\label{table:ground_photometry_358p}
\end{table*}

Calibrated photometry was further normalized to $r_h=\Delta=1$~au (producing reduced magnitudes) and then to a Solar phase angle of $\alpha=0^{\circ}$ (producing absolute magnitudes) to facilitate evaluation of the comet's intrinsic brightness evolution. Reduced magnitudes were computed by applying a correction of $-5\log (r_h\Delta)$ to the measured apparent magnitude, while absolute magnitudes were computed by determining the nucleus's reduced magnitude at the value of $\alpha$ at the time of observations using the IAU phase function parameters computed by \citet{hsieh2023_mbcnuclei}, subtracting the nucleus's reduced magnitude from the total measured reduced magnitude of the comet (i.e., leaving the reduced magnitude of the dust), correcting the reduced magnitude of the dust to $\alpha=0^{\circ}$ using the Schleicher-Marcus phase function\footnote{\url{https://asteroid.lowell.edu/comet/dustphase.html}} \citep[sometimes also referred to as the Halley-Marcus phase function;][]{schleicher2011_sw3,schleicher1998_halley,marcus2007_cometphasefunction}, and finally adding back the absolute magnitude of the nucleus.  We note that these calculations assumed that the dust coma is optically thin. 
These calculations and others described in subsequent sections use the {\tt uncertainties} {\tt python} package for the calculation and propagation of uncertainties\footnote{\url{http://pythonhosted.org/uncertainties/}}. 
Photometric results for 358P and 238P are shown in Tables~\ref{table:ground_photometry_358p} and \ref{table:ground_photometry_238p}.

\setlength{\tabcolsep}{5pt}
\setlength{\extrarowheight}{0em}
\begin{table*}[htb!]
\caption{$r'$-band Photometric Results for 238P$^a$}
\centering
\smallskip
\footnotesize
\begin{tabular}{ccccccccc}
\hline\hline
\multicolumn{1}{c}{UT Date}
 & \multicolumn{1}{c}{Telescope$^b$}
 & \multicolumn{1}{c}{$m_{4''}$}
 & \multicolumn{1}{c}{$m_{4''}(1,1,0)$}
 & \multicolumn{1}{c}{$Af\rho_{4''}$}
 & \multicolumn{1}{c}{$\rho_\mathrm{5000km}$}
 & \multicolumn{1}{c}{$m_\mathrm{5000km}$}
 & \multicolumn{1}{c}{$m_\mathrm{5000km}(1,1,0)$}
 & \multicolumn{1}{c}{$Af\rho_\mathrm{5000km}$}
 \\
\hline
2022 Aug 31 & Gemini-N    & 21.58$\pm$0.01 & 17.30$\pm$0.02 & 3.6$\pm$1.1 & 3.1 & 21.67$\pm$0.02 & 17.39$\pm$0.03 & 4.0$\pm$1.2 \\
2022 Sep 03 & Gemini-N    & 21.53$\pm$0.09 & 17.28$\pm$0.09 & 3.6$\pm$1.1 & 3.2 & 21.61$\pm$0.06 & 17.36$\pm$0.07 & 4.0$\pm$1.2\\
2022 Sep 08 & Gemini-N    & 20.97$\pm$0.06 & 16.78$\pm$0.06 & 3.9$\pm$1.1 & 3.3 & 21.45$\pm$0.06 & 17.26$\pm$0.06 & 4.2$\pm$1.3 \\
2022 Sep 22 & Gemini-N    & 21.46$\pm$0.07 & 17.44$\pm$0.07 & 3.5$\pm$1.0 & 3.5 & 21.50$\pm$0.07 & 17.49$\pm$0.07 & 3.7$\pm$1.1 \\
2022 Sep 22 & Gemini-S    & 21.42$\pm$0.01 & 17.40$\pm$0.02 & 3.7$\pm$1.0 & 3.5 & 21.44$\pm$0.01 & 17.43$\pm$0.02 & 3.9$\pm$1.1 \\
2022 Sep 30 & Palomar     & 21.44$\pm$0.08 & 17.53$\pm$0.08 & 3.4$\pm$1.0 & 3.7 & 21.47$\pm$0.06 & 17.56$\pm$0.06 & 3.5$\pm$1.0 \\
2022 Oct 01 & Gemini-N    & 21.34$\pm$0.01 & 17.46$\pm$0.02 & 3.6$\pm$1.0 & 3.7 & 21.39$\pm$0.01 & 17.50$\pm$0.02 & 3.7$\pm$1.1 \\
\hline
\hline
\multicolumn{9}{l}{$^a$ See Table~\ref{table:ground_photometry_358p} for explanations of column headings unless otherwise specified} \\
\multicolumn{9}{l}{$^b$ See Table~\ref{table:ground_observations_238p} for explanations of telescope designations.} \\
\end{tabular}
\label{table:ground_photometry_238p}
\end{table*}

To maximize signal-to-noise ratios to characterize the level of activity during each visit, we constructed composite images of the object for each night of data by shifting and aligning individual images on the object's photocenter using linear interpolation and then adding the images together (Figures~\ref{fig:358p_optical_images} and \ref{fig:238p_optical_images}).  Composite images were also constructed by shifting and aligning individual images on a sample of moderately bright field stars and adding the images together in order to enhance the visibility of faint stationary background sources near detections of the comet that could potentially contaminate photometric measurements of the comet.
When detections of the comet were deemed to be too close to background sources for photometry to be reliable (as determined from curve-of-growth analyses), photometric measurements of those detections were rejected.

\section{Results and Analysis}\label{section:results}

\subsection{Spectroscopic Analysis}

\subsubsection{H$_{2}O$ Production Rate Analysis\label{section:results_h2o_outgassing}}

Using the median-combined NIRSpec spectrum shown in Figure~\ref{fig:NIRSpec_extracted}, we derive 358P's water production rate from analysis of the $\nu_3$ H$_2$O emission band at 2.7~$\mu$m.  We first note that the emission band shape is not well-fit by a spectrum derived using a single dominant rotational temperature using the Planetary Spectrum Generator\footnote{\url{https://psg.gsfc.nasa.gov/}} \citep[PSG;][]{villanueva2018_psg}. In particular, a rotational temperature of $T_{\rm rot}=15$~K leads to a fit that matches the peak at 2.62~$\mu$m but overestimates the amplitude of the peak at 2.69~$\mu$m by 25$\%$, while a $T_{\rm rot}=30$~K characteristic temperature shifts the 2.62~$\mu$m emissions peak blueward, but matches the 2.69~$\mu$m emission profile in amplitude, but not shape.  This is similar to results found for 238P by \citet{kelley2023_jwst238p}. Whereas the four dominant emission lines underlying the two peaks have very different optical depths, the water column densities around 358P appear to be at least one order of magnitude too low to significantly alter their expected relative intensities \citep{crovisier1984_cometwater}. Rather, the need for multiple populations indicates that the coma was not in local thermodynamic equilibrium (LTE), likely due to low molecule-to-molecule and molecule-electron collisional rates in the inner coma resulting from 358P's low gas production rates.

Therefore, following \citet{kelley2023_jwst238p}, we obtain initial water production rates using a two-component non-LTE model incorporating spectra derived for rotational temperatures of $T_{\rm rot}=15$~K and $T_{\rm rot}=30$~K using the PSG for the observing conditions detailed in Table \ref{table:jwst_mbc_observations}, where we assume a constant gas outflow velocity of 515~m~s$^{-1}$. For reference, the best-fit model found by \citet{kelley2023_jwst238p} for 238P used rotational temperatures of $T_{\rm rot}=15$~K and $T_{\rm rot}=25$~K.  These best fit production rates are then passed to a Markov Chain Monte Carlo (MCMC) model, where they are used as initial estimates to explore the associated uncertainties with the {\tt emcee} {\tt python} package by simultaneously fitting a scaling constant and linear continuum function. 

\begin{figure}
    \centering
    \includegraphics[width=\linewidth]{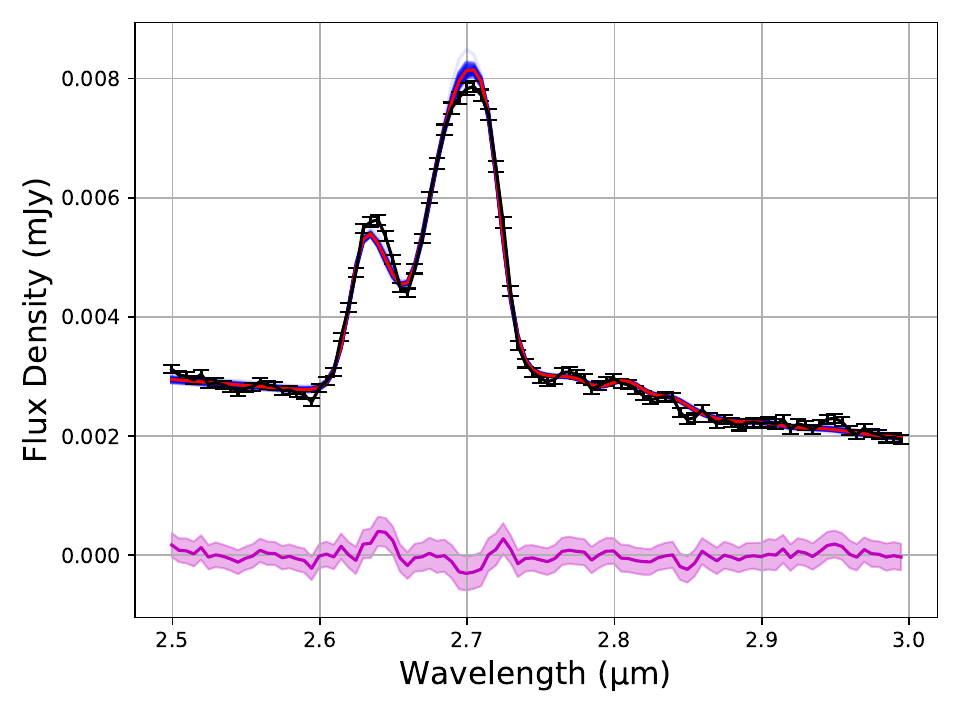}
    \caption{Comparison of measured H$_2$O emission with a modeled spectrum, where the measured H$_{2}$O 0$-$0 band emission at 2.69~$\mu$m for 358P is shown in black with error bars, and the modeled emission for a $0\farcs4$ beam for a Haser-like distribution of water outgassing with a total production rate of $Q_{\rm H_2O}=(5.0\pm0.2)\times10^{25}$~molecules~s$^{-1}$ split between two components with characteristic temperatures of 15~K and 30~K is shown in red. A small 0.003~\micron{} wavelength correction has been added to the JWST data, as is standard practice for PSG retrievals (G. Villanueva, priv. comm.).
    A representative sample of the walkers from the MCMC sampler are shown in blue and residuals and 3$\sigma$ errors are shown in magenta. Note that the excess residuals at 2.64~\micron{} and 2.72~\micron{} are just within the 3$\sigma$ error.
    }
    \label{fig:H2O_band}
\end{figure}

We measure an integrated H$_2$O flux of $(1.96\pm0.04)\times10^{-17}$~erg~cm$^{-1}$~s$^{-1}$, corresponding to an average column density of H$_2$O molecules within the measurement aperture of $(5.8\pm0.1)\times10^{15}$~m$^{-2}$.
The initial best fits from the PSG for each temperature return identical derived production rates of $Q_{\rm H_2O}=(4.6\pm0.2)\times10^{25}$~molecules~s$^{-1}$, but neither model satisfactorily matches the emission band shapes. Passing these best fits into {\tt emcee}, we retrieve a slightly larger production rate of $Q_{\rm H_2O}=(4.9\pm0.2)\times10^{25}$~molecules~s$^{-1}$ when simultaneously fitting the continuum with a single gas rotational temperature. We then use {\tt emcee} to fit both temperature emission profiles simultaneously to produce an improved fit that yields information on the production rate of each individual population. This combined fit is shown in Figure \ref{fig:H2O_band}. 
We find that the $T_{\rm rot}=15$~K population accounts for 40$\%$ of the production rate while the $T_{\rm rot}=30~K$ component is the remaining 60$\%$, and that the associated total $Q_{\rm H_2O}$ rises slightly to $Q_{\rm H_2O}=(5.0\pm0.2)\times10^{25}$~molecules~s$^{-1}$.
For reference, this production rate is $\sim5\times$ larger than the water production rate of $Q_{\rm H_2O}=(9.9\pm1.0)\times10^{24}$~molecules~s$^{-1}$ found for 238P \citep{kelley2023_jwst238p}.

\begin{figure*}
\centering
\includegraphics[width=0.49\linewidth]{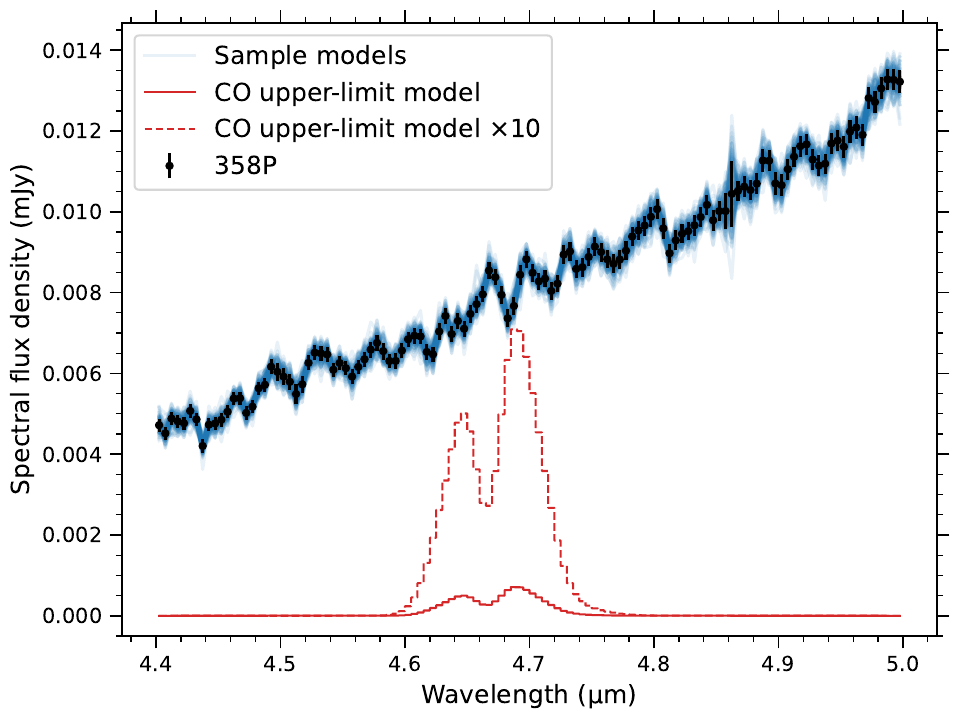}
\includegraphics[width=0.49\linewidth]{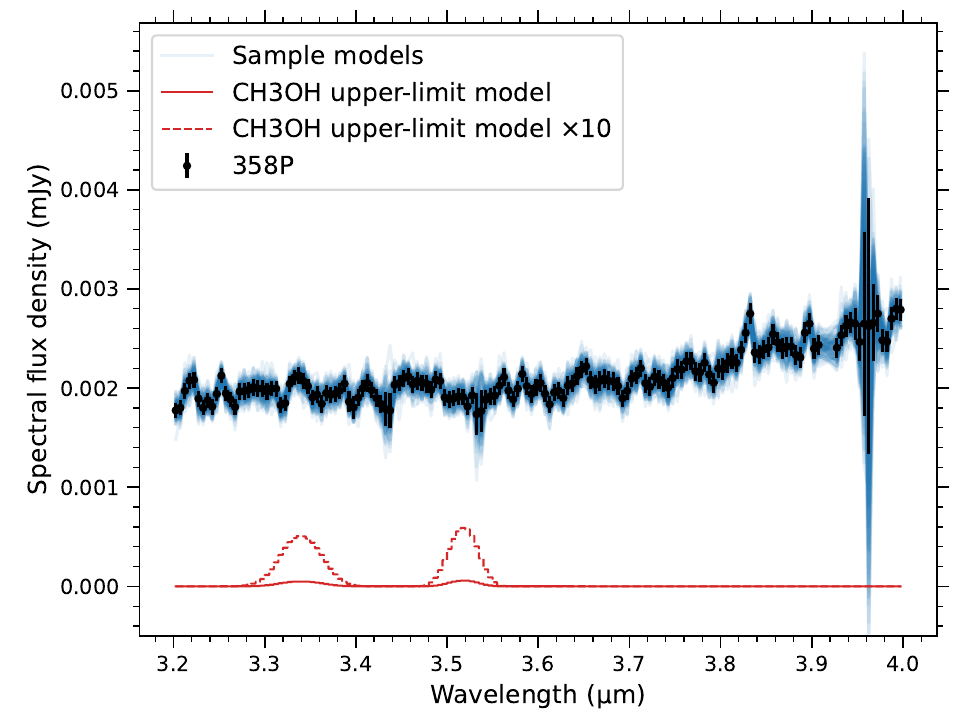}
\includegraphics[width=0.49\linewidth]{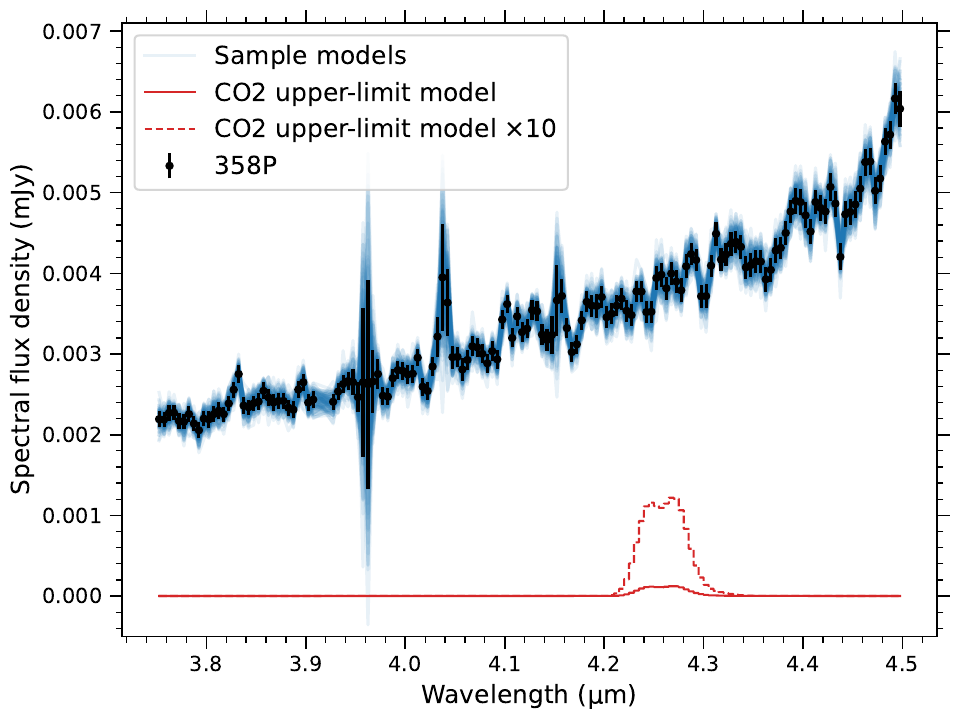}
\caption{Sample fits from the MCMC models for the CO (top left), CH$_3$OH (top right), and CO$_2$ (bottom) emission band regions for NIRSpec observations of 358P. Black points are observations and blue-to-gray lines indicate a subsample of the MCMC model fits to the data, showing the spread of those fits.  Also shown are model spectra for each gas with production rates equal to our 99.7-percentile upper limit, and 10 times that upper limit (thin solid and dashed red lines, respectively).}
\label{fig:gas_upper_limits}
\end{figure*}

We note that the fit is not perfect, and there is a slight under-prediction of the redward wing of the 2.62~$\mu$m peak as well as a slight overestimation of the blue wing of the 2.70~$\mu$m feature. We are unable to fully resolve these discrepancies with additional fine-tuning of the temperatures used in the model though, indicating that additional physical processes are relevant. We also note that our model does not allow the ortho-para ratio, or spin temperature, of the H$_2$O to be a free parameter; only the ortho symmetry is modeled. Measured cometary ortho-para ratios range from $\sim$2.5 \cite{faggi2018_C2017E4} to $\sim$3 for 67P/C-G \citep{cheng2022_orthopara67p} and $\sim$3.2 for Halley \citep{mumma1987_orthoparawater}. Indeed, the excess residuals shown in Figure \ref{fig:H2O_band} do not reflect an overestimate of the ortho-para ratio; if that were the case the residuals would be shifted to 2.66~\micron{} and 2.70~\micron{} \citep[see Fig.\ 2 of][]{cheng2022_orthopara67p}. OH prompt emissions or H$_{2}$O isotopologue emissions could also contribute to the residuals, but we note their presence will not change our derived H$_2$O production rate beyond the associated uncertainties.

\subsubsection{CO$_{2}$ and CH$_{3}$OH Production Rate Limits\label{section:results_hypervolatile_depletion}}

No clear emission due to any gas other than water is apparent in the median combined spectrum (Figure~\ref{fig:NIRSpec_extracted}). To obtain upper limits for 358P's CO$_{2}$, CO, and CH$_{3}$OH production rates, we implement MCMC characterizations of the production rates while simultaneously fitting the continuum background with second order polynomial models.  Also included in the fits are two parameter models that attempt to account for data covariance (we specifically used the exponential-squared kernel, $e^{-x^2/2}$, in the \texttt{George} package for Gaussian Process Regression; \citealt{ambikasaran2015_gaussianprocesses}).

\begin{figure*}
\centering
\includegraphics[width=\linewidth]{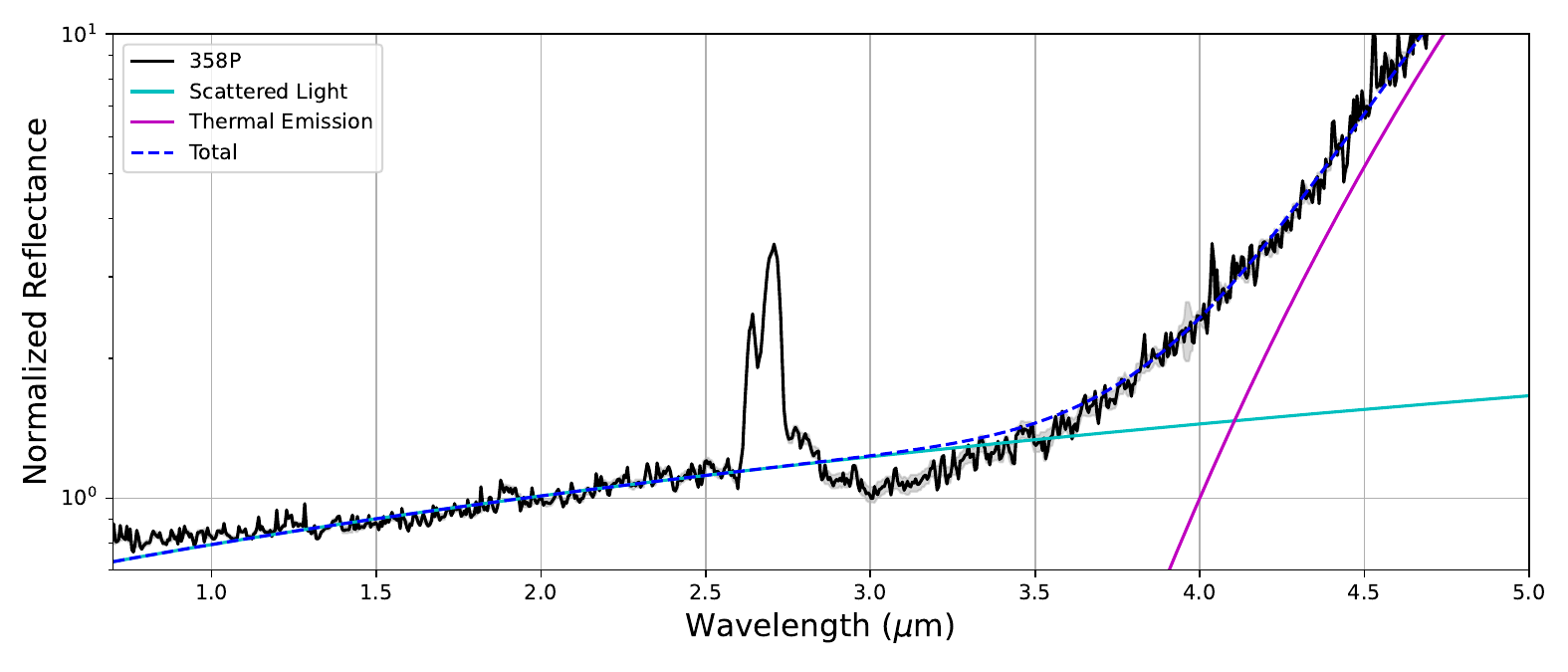}
\caption{Relative reflectance spectrum with 1$\sigma$ error in gray of 358P (black line) derived from the median-combined spectrum shown in Figure \ref{fig:NIRSpec_extracted} and the STScI CALSPEC Solar reference spectrum. Also shown are a linear reflectance (2.25$\pm$0.03 $\%$/100~nm) and blackbody curve fit (blue dashed line) to the observed reflectance, where the blackbody spectrum has a temperature of $T_{\rm BB}=200$~K and is normalized at 4~$\mu$m.}
\label{fig:358P_reflectance}
\end{figure*}

Models for the antisymmetric stretch in the $\nu_3$ band of CO$_2$ at 4.26~$\mu$m, the first excited state to the vibrational ground state of CO at 4.67~$\mu$m, and the $\nu_2$ and $\nu_3$ rovibrational bands of CH$_3$OH at 3.34 and 3.52~\micron{} are produced with the PSG for the NIRSpec observing conditions.
We assume a rotational temperature of $T_{\rm rot}=30$~K for CO and CH$_3$OH, the more abundant of the two temperature components (15~K and 30~K) used for H$_2$O.
For CO$_2$, a symmetric molecule with forbidden rotational transitions in the ground state, we set the rotational temperature to the ice sublimation temperature of 80~K.
After finding an initial best fit between the median extracted spectrum between 3.75~\micron\ and 4.5~\micron\ via the {\tt minimize} function from {\tt scipy.optimize}, we pass the model to the {\tt emcee} package, where 32 walkers are allowed to wander for 5000 iterations to fully characterize the errors in the fit.
We find 99.7 percentile (equivalent to $3\sigma$) upper limit CO and CH$_3$OH production rates of 
$Q_{\rm CO}=3.0\times10^{24}$~molecules~s$^{-1}$ and $Q_{\rm CH_3OH}=6.4\times10^{23}$~molecules~s$^{-1}$, or 6\% and 1\% with respect to water, respectively.
Meanwhile, assuming $T_{\rm rot}=80$~K as described above, we find a corresponding $3\sigma$ upper limit to the production rate of CO$_2$ of $Q_{\rm CO_2}=7.6\times10^{22}$~molecules~s$^{-1}$, or 0.2\% with respect to water.
For reference, this result is $\sim30$\% larger than an upper limit calculated using $T_{\rm rot}=30$~K, as assumed for CO and CH$_3$OH.
Sample model fits and the upper-limit gas models are shown in Figure~\ref{fig:gas_upper_limits}.

\subsubsection{Reflectance Spectroscopy\label{section:results_reflectance}}

To study 358P's near-infrared continuum, we divide the median-combined spectrum from Figure~\ref{fig:NIRSpec_extracted} by the STScI CALSPEC Solar reference spectrum to produce a relative reflectance spectrum (Figure~\ref{fig:358P_reflectance}).  
We find a mean linear spectral slope of $(2.25\pm0.03)$\% per 100~nm between 1.00~\micron\ and 2.55~\micron\ (normalized at 2.0~\micron), which is similar to the spectral slope of $(2.18\pm0.02)$\% found for 238P \citep{kelley2023_jwst238p}.

We model this reflectance spectrum by assuming that the scattered light has a constant spectral slope over the entire considered wavelength range and that the thermal contribution takes the form of a scaled blackbody spectrum.  Iterative fitting of the data is used to find a temperature for the blackbody spectrum of $T_{\rm BB}=200$~K.

The fits suggest that the thermal emission accounts for about 19\% of the spectrum at 3.7~\micron. The long-wavelength edge of the 3~\micron\ absorption feature is approximately 3.7~\micron, and therefore thermal emission may influence the shape of this feature. Assuming $r_n=0.3$~km and $p_V=0.05$ for 358P, and a nominal comet nucleus thermal model, we find that the nucleus accounted for 97\% of the spectral flux at 5.0~\micron.  At 2.0~\micron, reflected light from the nucleus accounts for 17\% of the spectral flux, assuming that the near-infrared colors of the nucleus and coma are similar.
For comparison, in the NIRSpec spectrum of 238P, the nucleus was found to account for $(98\pm37)\%$ of the spectral flux at 5.0~\micron, and $(21\pm8)\%$ of the spectral flux at 2.0~\micron\ \citep{kelley2023_jwst238p}.

In the spectra of both 238P and 358P, we see clear absorption around 3~\micron, where water ice is known to exhibit a broad absorption band.  Notably, however, we do not see absorption at 1.5~\micron\ or 2.0~\micron, where water ice is also known to exhibit absorption features, in the spectrum of either object.  Scenarios have been proposed to explain how absorption from water ice could be observed at 3~\micron\ but not at 1.5~\micron\ or 2.0~\micron\ \citep[e.g., for (24) Themis;][]{rivkin2010_themis}, but especially given that 3~\micron\ absorption can be produced by materials other than water ice \citep{takir2012_3micron,rivkin2022_3micron}, further investigation to determine the applicability of these scenarios to 238P and 358P would be useful.  
In the future, it will be interesting to see if similar absorption bands at 3~\micron\ continue to be found for other MBCs observed by \jwst{} (Section~\ref{section:discussion_future_work}).  Meanwhile,
comparison of the reflectance spectra of 238P and 358P, with a particular focus on the 3~\micron\ region, to those of other relevant reference objects observed by \jwst{} will also be useful for assessing the similarity or uniqueness of MBCs relative to other small body populations, which may in turn give insights into the plausibility of different formation scenarios for MBCs (Section~\ref{section:discussion_co2depletion}).

\subsection{Dust Characterization\label{section:results_dust}}

\subsubsection{Coma Properties and Photometry\label{section:results_coma}}

\begin{figure*}
    \centering
    \includegraphics[width=0.7\linewidth]{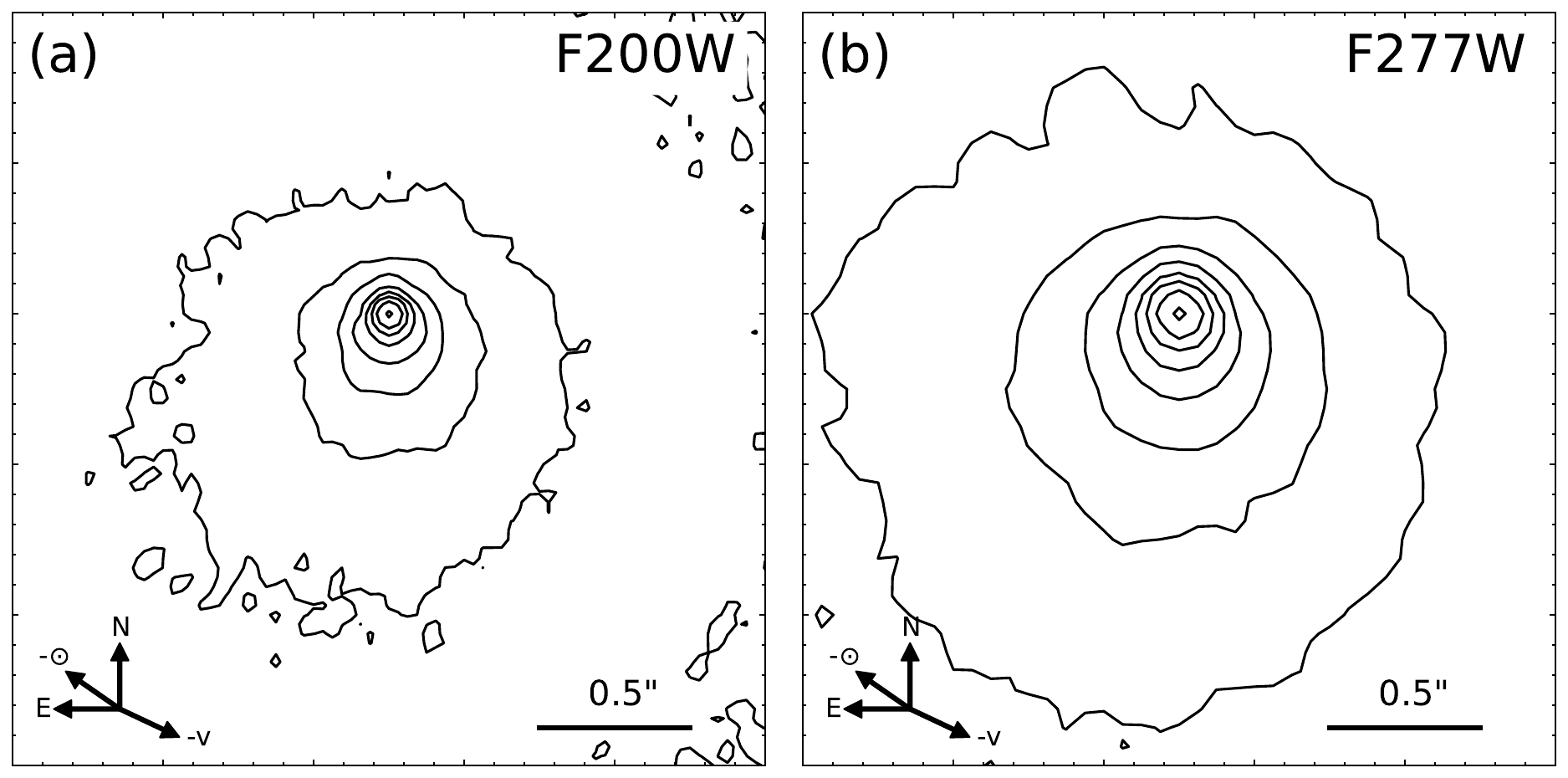}
    \caption{Contour plots (using 15 logarithmically spaced contour levels ranging from the peak value of each image --- 6.64~MJy/sr for F200W and 2.06~MJy/sr for F277W --- to the background level of $\sim0.01$~MJy/sr in each image) of the inner coma of 358P/PANSTARRS constructed from (a) F200W and (b) F277W NIRCam median composite images shown in Figures~\ref{fig:nircam_images}a and \ref{fig:nircam_images}b, respectively.  A $0\farcs5$ angular scale bar (760 km at the distance of the comet) is shown in each panel, where the orientations of the images and contour plots are the same.}
    \label{fig:nircam_contours}
\end{figure*}

Contour plots of the inner coma of 358P as imaged by NIRCam (Figure~\ref{fig:nircam_contours}) show that the coma deviates from circular symmetry beyond $\rho\sim0\farcs1$ from the photocenter in both the F200W and F277W median composite images.  Beyond this distance from the photocenter, the coma becomes visibly extended towards a position angle of ${\rm PA}\sim180^{\circ}$ East of North (i.e., almost directly South), which notably is not along the projected directions of either the comet's anti-Solar or negative heliocentric velocity vectors (Table~\ref{table:jwst_mbc_observations}; Figure~\ref{fig:nircam_images}). To investigate sources for this asymmetry we perform Monte Carlo modeling of the dust coma using an updated version of the model presented in \citet{kareta2023_46pcoma}, which we discuss further in Section \ref{section:results_dust_modeling}.
The coma appears slightly more centrally condensed in the F200W contour plot than in the F277W contour plot, but as this may simply be due to the difference in spatial resolution of observations in the two filters, we do not not regard it as significant.

Photometric measurements of the comet using $\rho=0\farcs3$ apertures result in measured average AB magnitudes of $22.25\pm0.01$~mag for our four F200W detections 
and $22.54\pm0.02$ for our four F277W detections, where uncertainties are based on the standard deviation of individual photometric points included in each average.  These results yield a ${\rm F200W}-{\rm F277W}$ color of $-0.29\pm0.02$~mag, compared to colors of ${\rm F200W-F277W}=-0.38\pm0.05$~mag and ${\rm F200W-F277W}=-0.39$~mag also measured using $\rho=0\farcs3$ photometry apertures for 238P from NIRCam and NIRSpec observations, respectively, obtained on UT 2022 September 8 \citep{kelley2023_jwst238p}.  Both the F200W and F277W bandpasses should be dominated by dust and nucleus continuum, but the latter bandpass also includes the 2.7~$\mu$m water emission band, and so, in principle, different ${\rm F200W}-{\rm F277W}$ colors could indicate differences in the relative contribution of water vapor to the observed coma.  However, in this case, as the ${\rm F200W}-{\rm F277W}$ colors for 238P and 358P are consistent within $2\sigma$, we do not find statistically significant differences in the relative compositions of the comae of the two objects from these data.

We do not detect 358P in data from our first night of optical ground-based observations (UT 2023 June 30; when the object was at $\nu=321.3^{\circ}$ and the nucleus had an expected $V$-band magnitude of $m_V=25.1\pm0.5$~mag).  Within the photometric apertures used for field stars on that night ($\rho=1\farcs8$), we find a point source detection limit of $m_r=23.8$~mag.  This detection limit aligns with the non-detection of 358P if the object was entirely inactive at the time. However, it also permits up to $\sim1.3$~mag of additional brightness due to potential activity, meaning we cannot rule out the possibility that the object was active in these data.

Meanwhile, photometry from observations on the night when 358P was first recovered (UT 2023 July 26) shows that its apparent magnitude was already $\sim1.5$~mag brighter than the expected $V$-band apparent magnitude of the inactive nucleus of $m_V=24.7$~mag, indicating that the comet was definitively active by the time it was at $\nu=328.8^{\circ}$.  Given that we cannot exclude 358P being active at the time of its non-detection on UT 2023 June 30 (at $\nu=321.3^{\circ}$), these results are consistent with the estimated activity start position at $\nu=(316\pm1)^{\circ}$ found by \citet{hsieh2018_358p} for 358P's 2017 active apparition.

Meanwhile, we find that the comet's near-nucleus absolute magnitude (as measured within photometry apertures with radii of $\rho=5000$~km at the distance of the comet) brightened by $\sim-3.5$~mag (a factor of $\sim25\times$) over about 150 days between UT 2023 July 26 ($\nu=328.8^{\circ}$) and UT 2023 December 21 ($\nu=11.9^{\circ}$) when the comet's near-nucleus coma reached its peak intrinsic brightness (or minimum absolute magnitude). After this peak, we find that the comet's near-nucleus absolute magnitude (measured using the same $\rho=5000$~km photometry apertures) faded by $\sim0.5$~mag (or a $\sim35$\% decline) over the next $\sim100$ days until UT 2024 March 02 ($\nu=32.6^{\circ}$).  For reference, we note that the comet's absolute magnitude peaked around $\nu\sim20^{\circ}$ in 2012-2013 \citep{hsieh2013_p2012t1}, while no observations are available to indicate when activity peaked during the comet's 2017 active apparition.

The long-lived activity of the comet with brightening and fading occurring over several months around each perihelion passage is consistent with observations of other MBCs, and is in sharp contrast, for example, with how quickly the disrupted asteroid (596) Scheila faded after it was impacted by another asteroid in 2010, where Scheila's brightness declined by 30\% in 8 days \citep{jewitt2011_scheila,bodewits2011_scheila}.  However, the active asteroid (6478) Gault, which is believed to be shedding mass due to rotational destabilization \citep[e.g.,][]{luu2021_gault,purdum2021_gault}, and not volatile sublimation, has been observed to exhibit long-lasting activity \citep{chandler2019_gault}, indicating that long activity duration is not reliable on its own as an indicator of sublimation-driven activity \citep[e.g.,][]{hsieh2012_scheila}.

\subsubsection{Photometric Dust Production Rate Characterization\label{section:results_dust_production_rates}}

In order to eventually place our derived water production rate for 358P (see Section~\ref{section:results_h2o_outgassing}) in context, we would like to quantitatively characterize the comet's dust production rate at the time of our \jwst{} observations as well as over the course of its full active apparition. 
The quantity $A(0^{\circ})f\rho$, hereafter $Af\rho$, is commonly used as a proxy for dust production rate that can be used to compare dust production activity levels derived from nucleus-subtracted and phase-function-corrected photometric measurements made of cometary coma, either for the same object or different objects, observed at different times and under different conditions \citep{ahearn1984_bowell}.  It is given by
\begin{equation}
    Af\rho = {(2r_h\Delta)^2\over\rho} 10^{0.4[m_{\odot}-m_d(r_h,\Delta,0)]}
\end{equation}
where $r_h$ is in au, $\Delta$ is in cm, $\rho$ is the physical radius in cm of the photometry aperture used to measure the magnitude of the comet at the distance of the comet, $m_{\odot}$ is the apparent magnitude of the Sun at $\Delta=1$~au in the same filter used to observe the comet, and $m_d(r_h,\Delta,0)$ is the phase-angle-corrected (to $\alpha=0^{\circ}$) apparent magnitude of the dust with the flux contribution of the nucleus subtracted from the measured total magnitude.

The parameter is nominally independent of $\rho$ for a spherically symmetric, steady-state coma with a line-of-sight column density that scales with $\rho^{-1}$ (which follows from a three-dimensional density profile that scales with $\rho^{-2}$) and no production or destruction of dust grains in the coma.
Notably, asymmetric dust ejection (e.g., in the form of jets or fans) should still produce an observed $\rho^{-1}$ radial surface brightness profile \citep{protopapa2014_103p}, preserving one of the key assumptions underlying the use of $Af\rho$ measurements as dust production rate proxies.  However, asymmetries over larger distance scales where coma morphology becomes dominated by radiation pressure effects can produce steeper surface brightness profiles \citep{jewitt1987_cometsbps,fink2012_afrho}, resulting in $Af\rho$ measurements that are not independent of $\rho$, but instead decrease with increasing $\rho$. Comet-to-comet and time-dependent variations in parameters like particle size distributions, expansion velocities, and specific physical natures of dust grains can also impact $Af\rho$'s exact correlation with dust production rates \citep{ahearn1995_ensemblecomets,fink2012_afrho,marschall2022_interceptor}, although for the purposes of our analysis here, we will neglect these higher-order effects.

We can obtain the approximate nucleocentric distance scale at which radiation pressure should begin to dominate the coma morphology using:
\begin{equation}
    X_{r} \sim {v_e^2r_h^2\over 2\beta_d g_{\odot}}
    \label{equation:turnaround_dist}
\end{equation}
\citep{jewitt1987_cometsbps}, where $X_{r}$ is the distance scale over which grains ejected sunward at a relative velocity of $v_e$ are turned around by Solar radiation pressure, $r_h$ is the heliocentric distance in au, $g_{\odot}=0.006$~m~s$^{-2}$ is the gravitational acceleration to the Sun at 1~au, and $\beta_d$ is the dimensionless ratio of Solar radiation pressure acceleration to the local Solar gravity (which varies approximately inversely with particle size, $a_d$, where $a_d$ in $\mu$m is approximately given by $1/\beta_d$).
Using an ejection velocity to grain size relation of $v_e(\beta_d)=v_0\beta_d^{1/2}$ \citep[e.g.,][]{moreno2013_p2012t1}, this relation simply becomes
\begin{equation}
    X_r \sim {v_0^2 r_h^2\over 2g_{\odot}}
\end{equation}
Using $v_0=40$~m~s$^{-1}$ as found by \citet{moreno2013_p2012t1} for their anisotropic emission model for 358P, and a typical heliocentric distance for our reported \jwst{} and ground-based observations of $r_h=2.4$~au, we find $X_{r}\sim800$~km.  This distance corresponds to projected angular distances on the sky of $0\farcs4<\theta_r<0\farcs7$ on the sky over the course of our ground-based observations of 358P ($1.43~{\rm au}<\Delta<2.86~{\rm au}$), and $\theta_r\sim0\farcs65$ during our \jwst{} observations ($\Delta=1.63$~au).  This result is relatively large compared to the circular symmetry scale of $\theta\sim0\farcs1$ for 358P's coma seen in NIRCam data discussed (Section~\ref{section:results_coma}; Figure~\ref{fig:nircam_contours}), although this discrepancy could be explained by asymmetric dust emission as suggested by \citet{moreno2013_p2012t1} and by our own dust modeling discussed below (Section~\ref{section:results_dust_modeling}).

Importantly, the sky-projected angular equivalents of $X_r$ during all but one set of our ground-based observations is smaller than the seeing disk (and in some cases, the pixel scale) on each night (see Tables~\ref{table:instrumentation} and \ref{table:ground_observations_358p}), meaning that any $Af\rho$ values derived from photometry from these observations will mostly represent radiation pressure-dominated dust, and will therefore be smaller than the ``true'' $Af\rho$ values that would be computed from the non-radiation pressure-dominated region of the coma alone.

\begin{figure}
    \centering
    \includegraphics[width=0.8\linewidth]{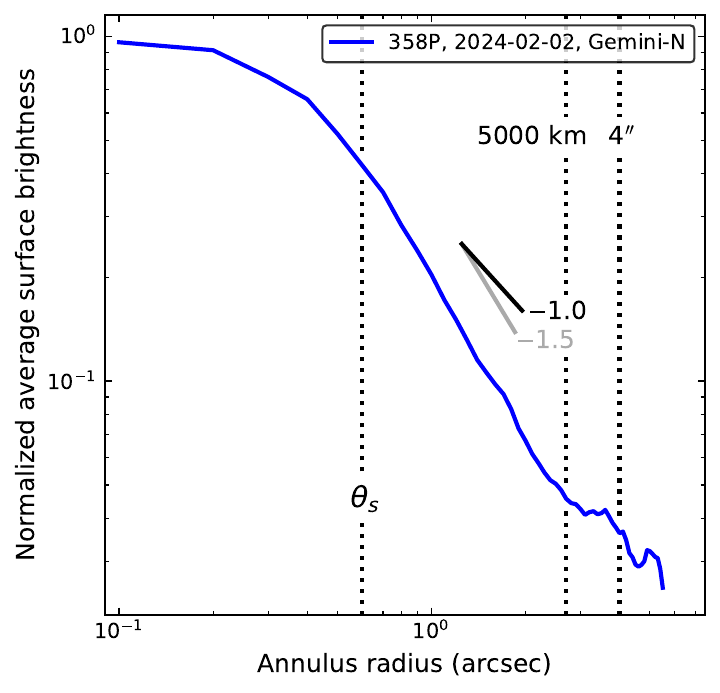}
    \caption{Normalized radial surface brightness profile of the non-nucleus-subtracted coma of 358P (blue line) measured from a composite image constructed from data obtained on UT 2024 February 2 by Gemini North, with diagonal lines indicating power law slopes of $-1$ (black line) and $-1.5$ (gray line) shown for reference.  Vertical dotted lines mark (from left to right) the FWHM seeing ($\theta_s$) from Table~\ref{table:ground_observations_358p}, $\rho=5000$~km, and $\rho=4''$.}
    \label{fig:358P_radial_profile}
\end{figure}

We find that the radial surface brightness profile (which includes both nucleus and coma flux) measured from our ground-based observations of 358P obtained on UT 2024 February 2 by Gemini North under $\theta_s=0\farcs6$ seeing conditions (our highest-quality ground-based observations within 30 days of our NIRSpec observations; see Table~\ref{table:ground_observations_358p}) follows a power law that is somewhat steeper than a $\rho^{-1}$ profile, consistent with the profile mainly comprising radiation pressure-dominated dust (Figure~\ref{fig:358P_radial_profile}).  This result is similar to profiles measured for 238P by \citet{hsieh2009_238p} as well as profiles measured for other radiation pressure-dominated comet comae by \citet{jewitt1987_cometsbps}.

\begin{figure}
    \centering
    \includegraphics[width=\linewidth]{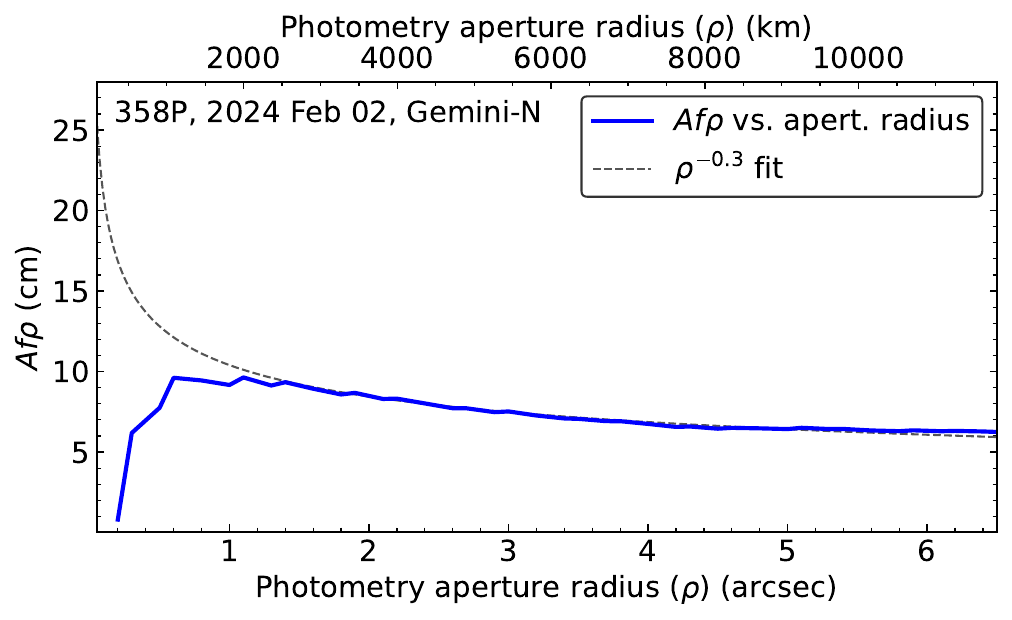}
    \caption{Plot of $Af\rho$ as a function of photometry aperture radius (solid blue line) in terms of arcseconds projected on the sky (bottom $x$-axis labels) and km at the distance of the comet (top $x$-axis labels) as measured for a composite image of 358P constructed from data obtained on UT 2024 February 02 by Gemini North, with a $\rho^{-0.3}$ power law (dashed black line) shown for reference.}
    \label{fig:358P_afr_vs_aprad_fit}
\end{figure}

Integrated photometry for a surface brightness profile proportional to $\rho^k$ scales with $\rho^{k+1}$
for $k\leq-1$
\citep[e.g.,][]{kelley2023_jwst238p},
and so we expect $Af\rho$ to scale similarly as a function of $\rho$ in our ground-based data.  To test this, we compute $Af\rho$ values from photometric measurements of our UT 2024 February 02 Gemini North observations of 358P made using a series of increasing photometry aperture radii in 1 pixel intervals (Figure~\ref{fig:358P_afr_vs_aprad_fit}), where we use $H_V=20.15\pm0.29$~mag and $G_V=0.18\pm0.28$ for the phase function parameters \citep{hsieh2023_mbcnuclei} used to compute the expected contribution of the nucleus to be subtracted from the total measured brightness of the comet, and the Schleicher-Marcus phase function (see Section~\ref{section:optical_data_reduction}) to calculate the expected apparent magnitude of the dust at $\alpha=0^{\circ}$.

We find that $Af\rho$ values for the coma computed in this way most closely matches a $\rho^{-0.3}$ power law (corresponding to a radial surface brightness profile power law slope of $k=-1.3$) over photometry aperture radii ranging from about $\rho=1\farcs5$ to $\rho=5\farcs0$.  Over this aperture radius range, $Af\rho$ values range from $Af\rho=9.3$~cm to $Af\rho=6.4$~cm.
Beyond this aperture range, photometry becomes increasingly impacted by sky noise and nearby background sources, while at smaller apertures, the subtraction of the entire expected nucleus contribution that we perform at all aperture sizes results in over-subtraction of the nucleus contribution (i.e., due to the use of aperture sizes that are comparable to or smaller than the effective size of the nucleus's point spread function), causing the downturn in $Af\rho$ values towards $\rho<1\farcs5$ seen in Figure~\ref{fig:358P_afr_vs_aprad_fit}.

Extrapolation to small $\rho$ using the power law used to fit the radiation-pressure-dominated region of 358P's coma in our ground-based data (e.g., Figure~\ref{fig:358P_afr_vs_aprad_fit}) would be a way to estimate the ``true'' $Af\rho$ value that would be measured in the non-radiation pressure-dominated region of the coma where $Af\rho$ is expected to be constant with $\rho$.
Past $Af\rho$ measurements of 358P and other MBCs were not made using this approach, however, and so to simplify comparisons to past observations, we opt here to simply measure $Af\rho$ values using similar aperture sizes as past analyses, while keeping in mind that those values represent lower limits to the ``true'' $Af\rho$ values that would be measured in the absence of radiation pressure.

For reference, analyzing observations obtained of 238P on UT 2022 September 08 with Gemini North using the approach described above, we find $Af\rho=3.9$~cm for a $4\farcs0$-radius aperture (equivalent to $\sim6000$~km at the distance of the comet).
\citet{kelley2023_jwst238p} estimated that 238P had $Af\rho=11.5$~cm at $\lambda=0.7$~$\mu$m \citep[where the $r'$ filter is centered at $\lambda_{\rm mid}=0.625$~$\mu$m with a width of 0.140~$\mu$m;][]{fukugita1996_sdss} from photometric measurements of NIRCam data using $0\farcs3$-radius photometry apertures.  Extrapolating this result for a radial surface brightness profile proportional to $\rho^{-1.5}$, they estimated it to be equivalent to $Af\rho=3$~cm using a $4\farcs0$-radius aperture, which is within the uncertainties of the value we derive from ground-based data using an actual $4\farcs0$-radius aperture.  

\begin{figure}
    \centering
    \includegraphics[width=\linewidth]{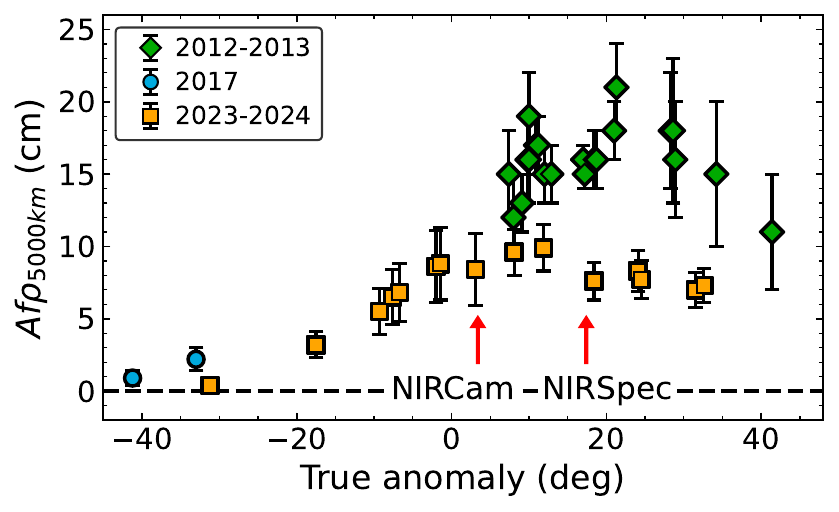}
    \caption{Plot of $Af\rho$ values reported for 358P during its 2012-2013 (green diamonds), 2017 (blue circles), and 2023-2024 (orange squares) active apparitions vs.\ true anomaly, where 2012-2013 and 2017 data are from \citet{hsieh2018_358p} and utilized $\rho=5\arcsec$ photometry apertures, and 2023-2024 data are from this work and utilized $\rho=5000$~km photometry apertures.  True anomaly positions at the time of NIRCam and NIRSpec observations by \jwst{} are marked with red arrows as labeled.}
    \label{fig:358p_afr_vs_nu}
\end{figure}

For internal consistency, we report $Af\rho$ measurements for photometry of our ground-based data using aperture radii of 5000~km at the distance of the comet, equivalent to $\rho=2\farcs4$ to $\rho=4\farcs8$ for our observations of 358P (Table~\ref{table:ground_observations_358p}).
In order to facilitate comparisons of these data to other data sets that may use different aperture size conventions, we also report uncertainties corresponding to the ranges of $Af\rho$ values measured using aperture radii of $\rho=(5000\pm2000)$~km.  Both sets of $Af\rho$ measurements obtained in these ways are listed in Tables~\ref{table:ground_photometry_358p} and \ref{table:ground_photometry_238p} for 358P and 238P, respectively.
We also plot the evolution of $Af\rho$ values for 358P as a function of true anomaly in Figure~\ref{fig:358p_afr_vs_nu}, illustrating the peak in near-nucleus brightness in 2023-2024 around $\nu\sim10^{\circ}$ and slow fading thereafter mentioned in Section~\ref{section:results_coma}, and where the true anomaly positions of our NIRCam and NIRSpec observations are also marked for reference.

In Figure~\ref{fig:358p_afr_vs_nu}, we also plot $Af\rho$ values reported by \citet{hsieh2018_358p} along with the new data reported in this work, which indicate decreased activity strength during 358P's 2023-2024 active apparition compared to its 2012-2013 active apparition, and even its 2017 active apparition.
Specifically, between $\nu=21.0^{\circ}$ and $\nu=28.6^{\circ}$ (i.e., a range covering the true anomaly of our UT 2024 February 02 Gemini North observations which occurred at $\nu=24.5^{\circ}$), 358P had an average value of $Af\rho\sim(19\pm4)$~cm in 2012-2013 based on photometry measured using $5''$-radius photometry apertures (equivalent to between 5500~km and 6200~km at the distance of the comet at the time of observations)  \citep{hsieh2018_358p}.  At the corresponding average physical photometry aperture size (5800~km) for our UT 2024 Feb 02 Gemini North data of $\rho=3\farcs2$, we measure $Af\rho=(7.3\pm1.0)$~cm for $\rho=(5800\pm2000)$~km (see Figure~\ref{fig:358P_afr_vs_aprad_fit}), suggesting that 358P's activity has weakened by about a factor of 2.5 since 2012.
This is corroborated by results found from dust modeling analyses (see Section~\ref{section:results_dust_modeling}).

\begin{figure*}
    \centering
    \includegraphics[width=0.485\linewidth]{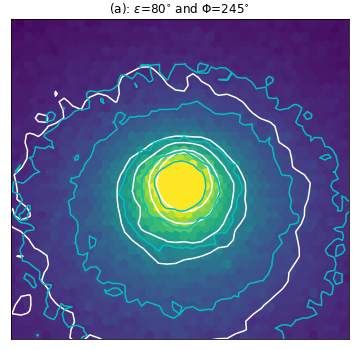}
    \includegraphics[width=0.50\linewidth]{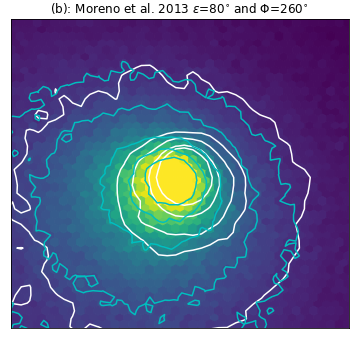}
    \caption{(a) Modeled image from a Monte Carlo dust model for $\mu$m-size grains accelerated by outgassing water from 358P,
    using an active South pole, $\varepsilon=80^{\circ}$, and $\Phi=245^{\circ}$, covering the inner $1500\times1500$~km of the coma as seen from \jwst{}. Intensities in the model image indicate column density, where representative contours are shown as cyan lines. NIRCam F200W contours from Figure~\ref{fig:nircam_contours} are shown in white for comparison. (b) The same as (a), but for $\Phi=260^{\circ}$ as found by \citet{moreno2013_p2012t1}.
    }
    \label{fig:dust_model_comp}
\end{figure*}

\subsubsection{Dynamical Dust Modeling}\label{section:results_dust_modeling}

In an effort to independently constrain the dust production rate for 358P and also gain insights into its activity in general, we perform a dynamical dust modeling analysis using our NIRCam images, focusing particularly on the south to southeast extension of the dust coma visible in those images (Figure~\ref{fig:nircam_contours}).  We use an updated version of the dynamical dust model described in \cite{kareta2023_46pcoma}, which takes into account the grain size distribution, radiation pressure, cometary gravity, and acceleration from outgassing volatiles. The updated version of the model takes the object's active area and obliquity into account. Unlike the models presented in \citet{moreno2009_sw1} and \citet{moreno2013_p2012t1}, which were designed to fit observed images, our model focuses on exploring the properties of dust particles within the near-nucleus coma. In our model, we use previously measured values of the radius and rotation rate of 358P's nucleus \citep{hsieh2023_mbcnuclei,agarwal2018_358pnucleus}, an estimated nucleus mass assuming a C-type asteroid-like density of $\rho_n=1300$~kg~m$^{-3}$ \citep{carry2012_astdensities}, the measured water production rate from our NIRSpec observations (assuming the rate to be the same during our NIRCam observations, which we expect to be sufficient for our purposes here),
and an assumed power law index for the dust size distribution of $-3.5$ \citep[the same value assumed by][]{moreno2013_p2012t1}.

Using the model parameters specified above, we are able to produce a synthetic image of the dust coma that is a reasonable match to NIRCam observations under two additional conditions: the dominant reflectance source in the F200W and F277W coma images are dust particles in the $\mu$m size range, and that the nucleus of 358P has an obliquity of $\varepsilon\sim80^{\circ}$ with a single active South polar region (latitudes $<-45^{\circ}$), similar to the best-fit asymmetric dust emission model found by \citet{moreno2013_p2012t1}. The resulting subsolar meridian at perihelion\footnote{The subsolar meridian at perihelion is defined by \citet{sekanina1981_rotation} as ``measured along the orbit from its ascending node on the equator (i.e. ``vernal equinox'') in the sense of the orbital motion''. To paraphrase, the subsolar meridian at perihelion is the angle between the rotational pole vector and the orbital velocity vector at perihelion when projected onto the plane of the orbit. Angles between $0^{\circ}$ and $180^{\circ}$ indicate that the northern hemisphere is facing the Sun at perihelion, while angles between $180^{\circ}$ and $360^{\circ}$ correspond to southern hemisphere illumination.} for 358P is then $\Phi=245^{\circ}$ for an active South pole (or $\Phi=70^{\circ}$ for an active North pole).  This corresponds to a right ascension and declination for the rotational pole of $\alpha\sim22.5^{\circ}$ and $\delta\sim+5.2^{\circ}$, respectively.

The resulting model output is shown in Figure \ref{fig:dust_model_comp}a. For reference, we compare this to output from our model where we use the values for obliquity ($\varepsilon=80^{\circ}$) and subsolar meridian at perihelion ($\Phi=260^{\circ}$) used by \citet{moreno2013_p2012t1} for their anisotropic model (Figure \ref{fig:dust_model_comp}b). 

Our model allows us to vary the velocities of individual particles as accelerated by both gas drag and radiation pressure, enabling the computation of more representative ranges of mass loss rates than were possible to compute prior to the direct measurement of water vapor production in MBCs.
We find that the velocity of $\mu$m-sized grains within the model is $\sim20-40$~m~s$^{-1}$. Larger grains are assumed to have velocities that scale as $a_d^{-0.5}$,
resulting in ${\bar a_d}\sim100$~$\mu$m
model grains \citep[the mass-weighted average size; e.g., see Equation~9 in][]{ishiguro2016_15p} moving at velocities between 0.5~m~s$^{-1}$ and 1.5~m~s$^{-1}$. For the range of projection angles relative to our \jwst{}-358P vector in our modeled active area, this leads to projected sky-plane velocities of up to 0.3~m~s$^{-1}$ for 100~$\mu$m grains.

To estimate the dust production rate at the time of our NIRCam observations, we convert the measured apparent F277W magnitude using a $0\farcs3$-radius aperture from Section~\ref{section:results_dust} to an absolute magnitude and then to a total surface area of $A_{\rm tot}=1.7\times10^{7}$~m$^{2}$ (assuming an albedo of $p=0.04$),
from which the scattering cross-section of the nucleus is subtracted to obtain a total dust cross section of $A_d=1.6\times10^{7}$~m$^{2}$. Using ${\bar a_d}=100$~$\mu$m, we then calculate a total coma mass within the $0\farcs3$-radius photometry aperture of $M_d=1.4\times10^{6}$ kg \citep[e.g., see Equation 8 in][]{ishiguro2016_15p},
assuming a dust grain density typical of CI and CM carbonaceous chondrites of $\rho_d=2500$~kg~m$^{-3}$ \citep{britt2002_astdensities_ast3}.
In the dynamical model, these particles have average on-sky velocities of $\sim0.3$~m~s$^{-1}$. At the projected angle relative to the Earth-358P vector ($\sim15^{\circ}-20^{\circ}$), they exit the aperture of 456~km in an average time of $\sim1.1\times10^{6}$~s. In the steady state assumption, that leads to an estimated mass loss rate of $\sim0.8$~kg~s$^{-1}$.  

This result is close to the dust production rate of $Q_{\rm dust}\sim0.7$~kg~s$^{-1}$ (assuming a maximum particle size of 1~cm) derived for the 2012-2013 active apparition by \citet{moreno2013_p2012t1} at an orbit position close to where our NIRSpec observations were taken (11 days after perihelion). Because of our different assumed particle densities (1000~kg~m$^{-3}$ vs.\ 2500~kg~m$^{-3}$), however, this result actually represents a decrease in the computed dust production rate between the two apparitions.  Adopting the same assumed particle density as \citet{moreno2013_p2012t1}, we find that the dust production rate we compute for 358P's 2023-2024 active apparition is $\sim2.5\times$ lower than that reported two orbits ago for the 2012-2013 active apparition, corroborating results found from our photometric analysis (Section~\ref{section:results_dust_production_rates}).

We do note that our modeling is based on NIRCam imaging from November 2023, when the water production rate, and therefore the corresponding mass loss rate, may have been slightly different at the time
compared to the dust mass loss rate at the time of our NIRSpec observations.
This is therefore a possible source of uncertainty when attempting to use this approach to compute the expected dust-to-gas production rate ratio at the time of our NIRSpec observations.

\subsection{Dust to Gas Ratios\label{section:results_gas_to_dust}}

Comparing our computed water production rate of $Q_{\rm H_2O}=(5.0\pm0.2)\times10^{25}$~molecules~s$^{-1}$,
or $Q_{\rm H_2O}\sim1.5$~kg~s$^{-1}$,
to our estimated dust production rate of $Q_{\rm dust}=0.8$~kg~s$^{-1}$ (Section~\ref{section:results_dust_modeling}),
we find 
$Q_{\rm dust}/Q_{\rm H_2O}\sim0.5$.
\citet{kelley2023_jwst238p} estimated a dust-to-gas production rate ratio of $Q_{\rm dust}/Q_{\rm H_2O}\approx0.3$ for 238P using a similar method, except that, like \citet{moreno2013_p2012t1}, their calculations also assumed a dust grain density of $\rho_d=1000$~kg~m$^{-3}$.  Using $\rho_d=2500$~kg~m$^{-3}$, this dust-to-gas ratio corresponds to $Q_{\rm dust}/Q_{\rm H_2O}\sim0.8$, which is in reasonable agreement with the value we find for 358P, considering the many assumptions incorporated into those results.

Performing a simple linear fit to $Af(\rho=5000~{\rm km})$ values that we find between UT 2023 December 08 and UT 2024 February 02 ($\sim30$ days before and after our NIRSpec observations), we estimate that 358P had $Af\rho=(8.6\pm1.5)$~cm at the time of our NIRSpec observations.  Combining this with our derived best-fit water production rate of $Q_{\rm H_2O}=(5.0\pm0.2)\times10^{25}$~molecules~s$^{-1}$, we obtain $\log_{10}(Af\rho/Q_{\rm H_2O})=-24.76\pm0.19$.
For comparison, we find $Af(\rho=5000~{\rm km})=(4.2\pm1.3)$~cm from ground-based observations of 238P on the same date as that target's NIRSpec observations (UT 2022 September 08; Table~\ref{table:ground_photometry_238p}), corresponding to $\log_{10}(Af\rho/Q_{\rm H_2O})=-24.37\pm0.14$, given its measured water production rate of $Q_{\rm H_2O}=(9.9\pm1.0)\times10^{24}$~molecules~s$^{-1}$ \citep{kelley2023_jwst238p}.

Neither value of $Af\rho/Q_{\rm H_2O}$ measured for 238P and 358P is an outlier among comparable measurements for other comets from the literature at similar heliocentric distances \citep[see][]{kelley2023_jwst238p}, indicating that despite the unusual depletion of hypervolatile species in MBC comae relative to other comets, dust-to-gas ratios are similar.  The relatively close agreement in the $Af\rho/Q_{\rm H_2O}$ ratios measured for two different MBCs furthermore suggests that $Af(\rho=5000)$~km could be usable as a proxy parameter for estimating water production rate at other times and for other objects for which NIRSpec observations are not available but ground-based observations are. This potentially enables the extension of limited \jwst{} data to the much larger collection of available ground-based data for MBCs, in terms of both the number of objects and time baselines covered.

\section{DISCUSSION}

\subsection{MBC Water Outgassing Detections\label{section:discussion_watervapor}}

Our detection of water vapor outgassing from 358P marks just the second such detection from a MBC after the detection of water vapor outgassing from 238P by \citet{kelley2023_jwst238p}.  Notably, prior to being observed by \jwst{}, both objects had already been identified as being likely to be exhibiting sublimation-driven activity based on their recurrent activity near perihelion and inactivity during intervening periods \citep{hsieh2011_238p,hsieh2018_358p,hsieh2018_238p288p}.  These results support the use of these characteristics as criteria for distinguishing MBCs from other active asteroids that exhibit dust emission driven by other processes (see Section~\ref{section:background}).
They furthermore confirm that MBCs, and therefore icy objects in general, may be widespread in the asteroid belt, given that 12 additional active asteroids (including both primary fragments of split comet 483P/PANSTARRS-A/B) at the time of this writing (see Appendix~\ref{section:appendix_recurrent_activity}) are also considered likely to be exhibiting sublimation-driven activity based on observations of recurrent activity near perihelion and inactivity during intervening periods \citep[where information on other active asteroids, including likely activity mechanisms whether or not recurrent activity has been confirmed, can be found in Table 1 in][]{jewitt2022_continuum_comets3}.
These objects span a range of physical and dynamical properties in the middle and outer main asteroid belt (i.e., with semimajor axes bounded by the 3A:1J and 2A:1J mean motion resonances with Jupiter, or $2.501~{\rm au}<a<3.277~{\rm au}$).

Using the ice sublimation model developed by \citet{cowan1979_cometsublimation} that is available at the NASA Planetary Data System's Small Bodies Node\footnote{\url{https://pds-smallbodies.astro.umd.edu/tools/ma-evap/index.shtml}}, we can compute the expected water ice sublimation rate, ${\dot m_w}$, from the surface of a sublimating gray-body at a given heliocentric distance. At $r_h=2.416$~au where 358P was observed by NIRSpec, we find ${\dot m_{w}}\sim3.5\times10^{20}$~molecules~s$^{-1}$~m$^{-2}$ in the non-rotating or pole-on case (which produces the maximum attainable temperature for an object), and ${\dot m_{w}}\sim0.97\times10^{20}$~molecules~s$^{-1}$~m$^{-2}$ in the zero-obliquity, fast-rotating case.  In the pole-on case (which our dust modeling analysis suggests is likely the most appropriate approximation for our situation; Section~\ref{section:results_dust_modeling}), our derived best-fit water production rate of $Q_{\rm H_2O}=(5.0\pm0.2)\times10^{25}$~molecules~s$^{-1}$ corresponds to an effective active area of $A_{\rm act}=0.14$~km$^2$.
This effective active area corresponds to a circular area $\sim200$~m in radius and an effective active fraction of $f_{\rm act}\sim0.13$ \citep[assuming a spherical nucleus with $r_n=0.3$~km for 358P;][]{hsieh2023_mbcnuclei}.
These derived estimates of $A_{\rm act}$ and $f_{\rm act}$ for 358P are both similar to the corresponding upper-limit values of $A_{\rm act}=0.11$~km$^{2}$ and $f_{\rm act}=0.15$ found for 238P in the cooler zero-obliquity, fast-rotating case \citep{kelley2023_jwst238p},
where these active fractions for both MBCs are well within the range of active fractions found for other comets \citep[e.g.,][]{samarasinha2013_c2012s1}.

We note that these effective active area and active fraction calculations are for hypothetical contiguous patches of exposed surface water ice. As such, in practice, assuming that active areas may have some degree of mantling that reduces effective sublimation rates, the total actual area of the comet's surface required to produce the observed gas production may be significantly larger than the effective area calculated here. The concept of active area itself also disregards the likely three-dimensional nature of the sublimation process in real icy objects \citep[e.g.,][]{nuth2020_asteroidvolatiles}.  As such, we encourage the reader to consider these calculations simply as a way to characterize relative activity levels as a function of nucleus size for the two \jwst{}-observed MBCs compared both to each other and to other comets.

\subsection{Interpretations of 358P's CO$_{2}$ Depletion\label{section:discussion_co2depletion}}

The coma abundance ratio of ${\rm CO}_2/{\rm H}_2{\rm O}\leq0.2$\% for 358P derived from our H$_2$O detection and CO$_2$ upper limit is even lower than the coma abundance ratio of ${\rm CO}_2/{\rm H}_2{\rm O}<0.7$\% measured for 238P (also by \jwst{})\footnote{The main cause for the lower relative CO$_2$ abundance limit in 358P compared to 238P is the $5\times$ higher water production rate of that comet. The spectral noise characteristics and the observing circumstances are otherwise similar.}, which was itself already an order of magnitude lower than previous spectroscopic measurements of other comets at similar heliocentric distances and a factor of three lower than any previous measurement for any comet ever \citep{kelley2023_jwst238p}.   Together, these measurements suggest that MBCs may be consistently strongly depleted in CO$_2$, and likely in other hypervolatile species or their daughter products (e.g., CN) as well.  This scenario is fully consistent with the numerous past non-detections of CN (see Section~\ref{section:background}) as well as thermal modeling by \citet{prialnik2009_mbaice} which suggested that hypervolatile species (specifically, HCN, NH$_3$, CO$_2$, and C$_2$H$_2$) should become completely depleted in main-belt asteroids on timescales far shorter than the age of the Solar system.

The question that then presents itself is whether this hypervolatile depletion is a consequence of the formation circumstances of MBCs, or their thermal evolution since formation.
If we presume that typical main-belt asteroids and classical comets from the outer Solar system are compositionally distinct due to different formation conditions, distinguishing these scenarios may require relatively simple analyses such as the acquisition of optical and near-infrared broadband colors or reflectance spectroscopy of inactive MBCs to evaluate whether they more closely resemble carbonaceous chondritic material characteristic of outer main-belt asteroids, or redder D-type-like colors or spectra found for comet nuclei from the outer Solar system \citep[e.g.,][]{lamy2004_cometnuclei_comets2, nuth2020_asteroidvolatiles}.  Unfortunately, the small sizes of most MBC nuclei \citep{hsieh2023_mbcnuclei}, and corresponding faintness ($m_V\sim23-26$~mag) when inactive makes it difficult to even acquire reliable broadband colors, much less spectra, although the Vera C.\ Rubin Observatory and its upcoming Legacy Survey of Space and Time \citep[LSST;][]{ivezic2019_lsst} may provide useful data for comparing MBC nucleus colors to other small bodies of similar sizes. Detailed cosmochemical analysis of MBC material would of course provide much more clarity regarding their formation circumstances \citep[e.g.,][]{hellmann2023_cchondrites}, although this would require a sample return spacecraft mission to a MBC.

Observations of MBC nuclei while they are inactive (e.g., at aphelion) with \jwst{} NIRSpec may provide useful comparisons to weakly active small cometary nuclei \citep[e.g., 157P/Tritton, 209P/LINEAR, 289P/Blanpain, and 460P/PANSTARRS;][]{jewitt2006_blanpain,pozuelos2014_shortperiodcomet_dustmodels,ye2016_209p,li2017_252p}.
In particular, such observations could focus on the 3-$\mu$m region \citep[where absorption features are seen in many outer main belt asteroids;][and references therein; also see Figure~\ref{fig:358P_reflectance}]{rivkin2022_3micron}, spectral slope, and any additional features that could help constrain formation regions. In particular, a search for evidence of ammoniated phyllosilicates or salts in the reflectance spectra of MBC nuclei would be timely and directly comparable to other small bodies, like 67P/Churyumov-Gerasimenko and the Trojan asteroids that will be observed by the \textit{Lucy} spacecraft \citep{raponi2016_67p,wong2024_lucy}. Recent work has shown that the presence of these minerals is likely indicative of formation on either side of the ammonia ice line \citep{rivkin2022_3micron}, with objects forming outside of the ammonia ice line sharing similarities with JFC nuclei. 

Hypervolatile depletion while forming within the ammonia ice line would indicate that MBCs formed in a location too warm for efficient CO/CO$_2$ accretion and retention.  Meanwhile, hypervolatile depletion while forming beyond the ammonia ice line could imply sequestration of CO/CO$_2$ into carbonates if ammoniated phyllosilicates signatures are present, a sign of aqueous alteration, or radiolytic formation of a H$_2$O/CO$_2$ ice matrix \citep{peeters2010_carbonicacid}. Carbonates have been detected on outer Solar system objects, but none as small as an MBC \citep{cartwright2024_jwst_ariel}. NIRSpec spectra of inactive MBC nucleus surfaces could then be further compared to those obtained for known icy outer Solar system objects as part of the DiSCo-TNOs large program on \jwst{} \citep[GO 2418;][]{pinillaalonso2023_discotnos,emery2024_jwsttnos} to assess the plausibility of an outer Solar system origin for MBCs.

Practically speaking, the strong depletion in hypervolatile species detected for both MBCs observed by \jwst{} to date means that ground-based searches for cometary sublimation products like CN that derive from hypervolatile species like HCN do not appear to be a viable means for confirming the sublimation-driven nature of MBC activity.  Furthermore, the limited availability of \jwst{} observing time means that it is likely infeasible to conduct individual searches for water vapor outgassing in all currently identified MBC candidates as well as in those that will be discovered in the future.  As such, it is useful to know that observations conducted to date support the viability and reliability of searching for recurrent active behavior as means for indirectly inferring the presence of sublimation-driven activity.

\subsection{Activity Evolution\label{section:discussion_activity_evolution}}

The $\sim2.5\times$ decline in 358P's activity strength between 2012-2013 and 2023-2024 provides useful data point for understanding long-term activity evolution in MBCs.  
If activity strength is assumed to always decrease monotonically over time (i.e., as a result of volatile depletion/recession and mantling), the rate of that activity attenuation is an important parameter for estimating a MBC's active lifetime, or the total time following an initial triggering event that a MBC remains active over any part of its orbit \citep[e.g.,][]{hsieh2009_htp,kossacki2012_259p}.  The average active lifetime for MBCs, in turn, is a key parameter for estimating the total number of currently active MBCs expected given an assumed rate of activity triggering and the total number of ``potential'' MBCs (i.e., icy asteroids with the physical and dynamical properties needed to have the potential to become active), which can then be compared against population estimates derived from observational surveys \citep[e.g.,][]{hsieh2009_htp,gilbert2009_cfhtmbcs,gilbert2010_cfhtmbcs,sonnett2011_cfhtmbcs,waszczak2013_ptfmbcs,hsieh2015_ps1mbcs} to assess the plausibility of those assumptions and refine them as needed.

Regarding other MBCs for which activity evolution has been explicitly studied, however, \citet{hsieh2018_238p288p} found that 238P's activity strength over similar true anomaly ranges declined by a factor of $\sim2$ between its 2010-2011 and 2016-2017 active apparitions, while activity strength may have actually increased by a factor of $\sim1.5$ between 2000 and 2016-2017 for 288P (although the authors note that they consider this a tentative result requiring additional confirmation).  \citet{hsieh2010_133p} reported that the activity strength of 133P/Elst-Pizarro appeared to have stayed approximately constant between its 2002 and 2007 active apparitions, although that work did not compare observations obtained over similar true anomaly ranges.

Further work is clearly needed to better understand the diversity of activity evolution behavior exhibited by MBCs.  In particular, if increasing activity strength can be confirmed for other objects, it could point to more complex physical processes affecting activity evolution than simple volatile depletion.
Such processes could include the exposure of new volatile reservoirs during active apparitions, perhaps via processes like sublimation-induced sinkhole collapses or landslides \citep[e.g.,][]{vincent2015_cometsinkholes,pajola2017_cliffcollapse,wesolowski2020_cometoutbursts}, rather than the steady depletion of the same subsurface ice reservoirs over the course of multiple active apparitions. We note that these processes could either cause activity to weaken over time by burying subsurface ice under increasing amounts of inert material, essentially accelerating the mantling process, or strengthen over time by excavating fresh volatile reservoirs.
Pole precession, perhaps due to outgassing \citep[e.g.,][]{kramer2019_67poutgassing}, could also cause changes in seasonal heating conditions that could increase or decrease heating of localized active areas (although 358P's weak activity presumably makes this scenario relatively unlikely). 
Lastly, sublimation of volatiles from deeper subsurface layers and refreezing at shallower depths before they are lost to space \citep[e.g.,][]{luspaykuti2022_cometoxygen,nuth2020_asteroidvolatiles} could also cause more complex activity evolution behavior.

It is unclear at this time whether the slight shift in the timing of 358P's activity peak to earlier in its orbit in 2023-2024 as compared to 2012-2013 noted in Section~\ref{section:results_coma} is physically significant.
If the delay in the activity peak past perihelion is due to the finite time required for a thermal wave to penetrate the comet's surface and reach subsurface ice reservoirs \citep[e.g.,][]{hsieh2011_238p}, we might nominally expect the activity peak to shift later, not earlier, in the orbit during subsequent orbits as ice depletion from sublimation causes the ice front to recede deeper below the comet surface with each active apparition.  This means that the trend seen at the moment for 358P of activity shifting earlier rather than later in the orbit, if real, could indicate the action of more complex physical processes controlling activity strength as a function of orbit position like those suggested above for explaining increasing activity strength over time.

Unfortunately, 358P's Solar elongation of $<40^{\circ}$ between $\nu\sim340^{\circ}$ and $\nu\sim25^{\circ}$ during its 2017-2018 active apparition means that no data are available to constrain the activity peak during that apparition.
358P will have a Solar elongation of  $>40^{\circ}$ after its next perihelion passage between $\nu\sim5^{\circ}$ and $\nu\sim90^{\circ}$ in mid-2029 (although unfortunately will not be observable just before that from $\nu\sim315^{\circ}$ and $\nu\sim5^{\circ}$, meaning that the onset time of activity is unlikely to be observable for that perihelion approach), and so it should be possible to observationally constrain its activity peak during that apparition to see if the apparent trend of the peak shifting earlier in the orbit over time continues.

In the meantime, thermal modeling analyses to assess the physical plausibility of subsurface ice front recession causing activity peaks to shift would be valuable.
Additionally, dust modeling analyses to determine the necessity at all of invoking thermal wave propagation time to to explain post-perihelion coma brightness peaks (e.g., instead of such peaks simply being due to large dust particles ``piling up'' inside photometry apertures over time) would also be extremely useful.

\subsection{Significance of 358P's Obliquity\label{section:discussion_obliquity}}

Thermal modeling has shown that objects with extreme obliquity ($\varepsilon\sim0^{\circ}$ or $\varepsilon\sim180^{\circ}$) are better able to preserve shallow subsurface ice near their poles \citep{schorghofer2008_mbaice,schorghofer2018_asteroidiceloss,schorghofer2020_icepreservation}, which is interesting in the context of the potential prevalence of dormant MBCs (i.e., inactive but icy asteroids) given the clustering of main-belt asteroid spin poles toward extreme obliquities found from Gaia data \citep{durech2023_gaiaspinstates}.
However, there is an interesting contradiction in that the same lower temperatures at the polar regions of extreme obliquity objects that help to preserve subsurface ice over long timescales will also limit both the likelihood and strength of any sublimation of that subsurface material that does occur.  In other words, while extreme obliquity is more thermally favorable for retaining shallow subsurface ice in asteroids until the present day, mid-range obliquity ($\varepsilon\sim90^{\circ}$) is actually more favorable for producing observable sublimation-driven activity from active MBCs \citep[][]{zhang2023_mbcactivity_acm}.

These issues are particularly relevant in the case of 358P, given its mid-range obliquity, initially suggested by \citet{moreno2013_p2012t1} through dust modeling analysis and further corroborated by our own dust modeling results. It has also been suggested that 133P has a mid-range obliquity \citep{kaluna2011_133pactivity}, although this result has yet to be conclusively confirmed. A mid-range obliquity ($50^{\circ}<\varepsilon<75^{\circ}$) has also been inferred for 176P/LINEAR \citep{hsieh2011_176p}.

Resolving the apparent inherent contradiction in the favorability of different obliquity conditions for the dormant and active phases of a MBC's life cycle could be accomplished in various ways.  One possibility is that MBC activity could actually be triggered by changes in obliquity, perhaps due to impacts \citep[e.g.,][]{paolicchi2002_asteroidcollisions}, radiative forces \citep[e.g.,][]{vokrouhlicky2002_yorp}, or gravitational interactions with other bodies \citep[e.g.,][]{scheeres2000_asteroidspin}.  This explanation could account for both the preference for extreme obliquity for long-term preservation of ice until the present day and the preference for mid-range obliquity for current active sublimation.  

Interestingly, results from NASA's Double Asteroid Redirection Test \citep[DART;][]{rivkin2021_dart,daly2023_dart} mission indicate that, along with changing an object's spin state, an impact could actually cause global deformation of the entire object \citep{richardson2024_dart}, depending on its physical properties, internal structure, and impact circumstances.  Such deformation could potentially expose subsurface ice on a global scale across the object's surface and thus further help to initiate activity.  In the case of DART, the impact target, Dimorphos, has an effective radius of $r_n=75$~m \citep{daly2024_dimorphos}, making it about 1/4 the size of 358P (or 1/64 of the mass, assuming similar bulk densities).  Meanwhile, the DART spacecraft had an impact velocity of $v\sim6$~km~s$^{-1}$ \citep[similar to typical relative encounter velocities between main-belt asteroids; e.g.,][]{farinella1992_astcollisionrates} and a mass at impact of $m_{sp}=580$~kg \citep{cheng2023_dart}, equivalent to a  spherical asteroid impactor with $r_n\sim0.4$~m, assuming a bulk density of $\rho_n=2500$~kg~m$^{-3}$.  Assuming that a similar mass ratio is needed to produce a similar effect (for approximately similar relative velocities), this suggests that an impactor $64\times$ the size of a DART-equivalent asteroid or larger \citep[i.e., $r_n\gtrsim25$~m, comparable to the upper range of impactor sizes previously proposed as being responsible for triggering MBC activity;][]{hsieh2009_htp,capria2012_mbcactivity} and a similar impact velocity could, in principle, have similar global deformation effects on 358P \citep[as well as other similarly sized MBCs; see][]{hsieh2023_mbcnuclei} as DART had on Dimorphos.

Another explanation could be that, as suggested by \citet{hsieh2018_activeastfamilies}, most or all MBCs could be relatively young fragments of recent family-forming catastrophic disruptions of larger parent asteroids.  In this scenario, the parent asteroids could be the bodies that previously had extreme obliquity and thus were able to preserve substantial ice near their poles, while the present-day family members could have effectively random obliquities, but are young enough that subsurface ice has yet to recede significantly, even in mid-range obliquity objects.

Finally, \citet{schorghofer2020_icepreservation} found that for small to moderate thermal inertias, rotational poles can still have the coldest temperature averaged over an entire orbit (i.e., the temperature most representative of the subsurface) of any other location on a body.  Therefore, it is possible that extreme obliquity is not a necessary condition for preserving volatile material on MBCs at all.

While the small sizes of most MBC nuclei \citep[see][]{hsieh2023_mbcnuclei} mean that they are typically extremely faint when inactive and thus difficult to measure rotational properties for, efforts to determine the obliquities for more MBCs, and thus the overall obliquity distribution of the population, would be extremely useful for assessing the significance of obliquity for MBC activity.  Such efforts could include lightcurve inversion analyses \citep{kaasalainen2001_lightcurveinversion1,kaasalainen2001_lightcurveinversion2,durech2023_gaiaspinstates} of the few MBCs that are bright enough for meaningful lightcurve data to be obtained while they are inactive. Dust modeling analyses like the ones performed as part of this work and by \citet{moreno2013_p2012t1} for 358P that could also be able to constrain obliquities from dust emission dynamics.  Additional theoretical consideration of the expected stability of obliquities of main-belt asteroids and the responses of loose rubble piles to impacts could also shed light on the plausibility of obliquity changes and/or impact-induced object deformation being triggers of MBC activity.

\setlength{\extrarowheight}{0em}
\begin{table*}[htb!]
\caption{Completed \jwst{} Observations of MBCs}
\centering
\smallskip
\footnotesize
\begin{tabular}{cccccrcccc}
\hline\hline
\multicolumn{1}{c}{UT Date$^a$}
 & \multicolumn{1}{c}{Target}
 & \multicolumn{1}{c}{Instrument$^b$}
 & \multicolumn{1}{c}{$a$$^c$}
 & \multicolumn{1}{c}{$e$$^d$}
 & \multicolumn{1}{c}{$i$$^e$}
 & \multicolumn{1}{c}{$H_V$$^f$}
 & \multicolumn{1}{c}{$r_n$$^g$}
 & \multicolumn{1}{c}{$r_h$$^h$}
 & \multicolumn{1}{c}{$\nu$$^i$}
 \\
\hline
2022 Sep 08 & 238P & NC, NS & 3.166 & 0.252 & 1.264 & 20.5$\pm$0.1 & 0.24$\pm$0.07 & 2.428 & 28.3  \\
2023 Nov 22 & 358P & NC & 3.147 & 0.239 & 11.060 & 20.2$\pm$0.3 & 0.29$\pm$0.10 & 2.396 &  ~\,3.4 \\
2024 Jan 08-09 & 358P & NS & 3.147 & 0.239 & 11.060 & 20.2$\pm$0.3 & 0.29$\pm$0.10 & 2.416 & 17.4 \\
2024 Jun 12 & 133P & NS & 3.164 & 0.156 & 1.390 & 15.88$\pm$0.04 & 2.0$\pm$0.6 & 2.674 & ~\,7.9 \\
2024 Sep 20 & 457P & NC, NS & 2.645 & 0.120 & 5.225 & 19.25$\pm$0.13 & 0.42$\pm$0.03 & 2.335 & ~\,9.0 \\
2024 Oct 14 & 133P & NC, NS & 3.164 & 0.156 & 1.390 & 15.88$\pm$0.04 & 2.0$\pm$0.6 & 2.747 & 37.4 \\
2024 Oct 28 & 133P & NC & 3.164 & 0.156 & 1.390 & 15.88$\pm$0.04 & 2.0$\pm$0.6 & 2.760 & 40.6 \\
\hline
\hline
\multicolumn{10}{l}{$^a$ UT date of completed observations, or UT date range for planned observations, as indicated.} \\
\multicolumn{10}{l}{$^b$ Instrument used (NC: NIRCam; NS: NIRSpec).} \\
\multicolumn{10}{l}{$^c$ Semimajor axis, in au.} \\
\multicolumn{10}{l}{$^d$ Eccentricity.} \\
\multicolumn{10}{l}{$^e$ Inclination, in degrees.} \\
\multicolumn{10}{l}{$^f$ Absolute $V$-band magnitude, from \citet{hsieh2023_mbcnuclei} and \citet{kim2022_p2020o1}.} \\
\multicolumn{10}{l}{$^g$ Nucleus radius, in km, from \citet{hsieh2023_mbcnuclei} and \citet{kim2022_p2020o1}.} \\
\multicolumn{10}{l}{$^h$ Heliocentric distance at time of observations, in au.} \\
\multicolumn{10}{l}{$^i$ True anomaly at time of observations, in degrees.} \\
\end{tabular}
\label{table:past_future_jwst_obs}
\end{table*}

\subsection{Future Work\label{section:discussion_future_work}}

Now that water vapor has been clearly detected in both MBCs that have been observed by \jwst{} thus far, a next obvious step is to consider the factors that affect or control the properties of that water vapor production. 
For instance, as noted in Section~\ref{section:results_h2o_outgassing}, we find a water production rate for 358P in this work that was $5\times$ higher than the rate found for 238P.
This is interesting given that both objects are similar in size ($r_n\sim0.25$~km for 238P and $r_n\sim0.30$~km for 358P) and were observed at similar heliocentric distances and true anomalies ($r_h=2.428$~au and $\nu=28.3^{\circ}$ for 238P, and $r_h=2.416$~au and $\nu=17.4^{\circ}$ for 358P).  
Using the ice sublimation modeling tool used in Section~\ref{section:discussion_watervapor}, we find that, based on thermal considerations alone, the difference in $r_h$ between the 238P and 358P NIRSpec observations should have only resulted in a 4\% higher water production rate
in the zero-obliquity fast-rotator case, and 2\% higher water production rate
in the non-rotating or pole-on case, assuming the same obliquity for each object.  The difference in the measured water production rates of 238P and 358P therefore must be mostly due to other factors.
Assessing the effects of different properties on activity strength will require continuing \jwst{} observations of additional MBCs with a range of physical properties under different observing circumstances to characterize water production rates, dust-to-gas ratios, and coma compositions, and continuing efforts to characterize the physical properties of those MBCs like nucleus sizes \citep[e.g.,][]{hsieh2023_mbcnuclei} and rotational states (Section~\ref{section:discussion_obliquity}) via analyses of ground-based observations.

At the time of the writing of this manuscript, five sets of \jwst{} observations of MBCs have been completed as part of \jwst{} Programs 1252, 4250, and 5551 in Cycles 1, 2, and 3, respectively.  Table~\ref{table:past_future_jwst_obs} lists the observing circumstances of these observations and information on the targets themselves.  Four unique targets (133P, 238P, 358P, and 457P/PANSTARRS) have been observed, where 133P was observed twice at different times, and therefore at different points in its orbit and at different heliocentric distances.

These targets span a range of dynamical and physical properties, with 457P notably being the only target with $a<3$~au, and 133P's nucleus being about an order of magnitude larger than those of the other targets.  This collection of observations will enable initial exploration of the level of variation of water vapor outgassing rates, $Q_{\rm dust}/Q_{\rm H_2O}$ ratios, CO$_2$/H$_2$O coma abundance ratios, and other outgassing properties among different objects, as well as the dependence of these properties on parameters such as object size, orbital elements, heliocentric distance, and orbit position.  
Fully understanding the dependence of MBC outgassing properties on these various parameters, however, will still require additional \jwst{} observations of MBCs to more thoroughly sample the available parameter space of physical and dynamical properties.

Otherwise, \jwst{} observations of other active objects in the Solar system will be extremely useful for drawing distinctions (e.g., comparing ${\rm F200W}-{\rm F277W}$ colors) between MBCs and other comet populations.  Such information should help to identify properties that are unique to MBCs, and thus could be characteristic of their unique formation region or subsequent thermal evolution relative to other comets, and which properties are more universal among all types of comets, and thus may be intrinsic properties of sublimation-driven dust emission in general.
Additionally, our non-detections of CO$_2$ and CO require us to assume their rotational temperatures in order to derive their upper limit production rates (in the case of this work, by assuming the average of two rotational temperatures required to fit the detected water emission feature at 2.7~\micron{}), leading to uncertainty in those calculated limits (Section~\ref{section:results_hypervolatile_depletion}).
\jwst{} observations of classical but low-activity comets --- i.e., comets that are not extremely depleted in hypervolatiles --- could enable the exploration of the rotational temperatures of major cometary volatiles in low-activity comets (i.e., with low-density near-nucleus comae) to discern whether they indicate direct sublimation of those volatiles or perhaps their release as the surrounding water ice matrix sublimates away, or could be influenced by other factors \citep[e.g.,][]{radeva2013_2P,saki2024_cometvolatiles}.

Finally, as discussed in Section~\ref{section:discussion_watervapor}, the confirmation of recurrent activity near perihelion and inactivity during intervening periods by ground-based facilities as an apparently reliable indicator of sublimation-driven activity underscores the importance of continuing to obtain such observations to identify other objects that are likely to be exhibiting volatile sublimation.  Such observations will be crucial both for identifying potential future targets for detailed \jwst{} studies, and building up the known population of sublimating main-belt objects from ground-based observations alone (i.e., without the need to obtain \jwst{} confirmation of volatile outgassing for every individual candidate member of the population).
Given results suggesting that reasonable first-order estimates of the water production rates of MBCs can be obtained from ground-based optical photometry data alone (Section~\ref{section:results_gas_to_dust}), optical monitoring during the active periods of MBCs could also be used to characterize the evolution of water sublimation rates for MBCs along their orbits and to compare relative water production rates among a larger portion of the MBC population than could feasibly be directly observed by \jwst{}.

\section{Summary}

In this work, we present the following key findings:
\begin{enumerate}
\item{We report a detection of water vapor outgassing in main-belt comet 358P/PANSTARRS in spectroscopic observations obtained on UT 2024 January 8-9 by \jwst{}'s NIRSpec instrument, where we find a best-fit water production rate of $Q_{\rm H_2O}=(5.0\pm0.2)\times10^{25}$~molecules~s$^{-1}$.  Along with an earlier water vapor detection made by NIRSpec for another main-belt comet (238P/Read), these results support the proposition that observations of recurrent activity near perihelion and inactivity during intervening periods is a viable and reliable means for identifying sublimation-driven activity in active asteroids.}

\item{As in previous NIRSpec observations of 238P, we find a remarkable lack of hypervolatile species, where we find $3\sigma$ upper limits of $Q_{\rm CO_2}=7.6\times10^{22}$~molecules~s$^{-1}$}, Q$_{\rm CH_3OH}=6.4\times10^{23}$~molecules~s$^{-1}$, and $Q_{\rm CO}=3.0\times10^{24}$~molecules~s$^{-1}$. These measurements yield a CO$_2$/H$_{2}$O coma abundance ratio for 358P of $Q_{\rm CO_2}/Q_{\rm H_{2}O}<0.2$\%, which is lower than the ratio of $Q_{\rm CO_2}/Q_{\rm H_{2}O}<0.7$\% measured for 238P from previous \jwst{} observations, which was itself already an order of magnitude lower than previous spectroscopic measurements of other comets at similar heliocentric distances and a factor of three lower than any previous telescopic measurement for any comet ever.

\item{Images obtained with \jwst{}'s NIRCam instrument show a clearly visible asymmetric coma that deviates from circular symmetry beyond radii of $\sim0\farcs1$ from the comet's photocenter in both F200W and F277W median composite images. The coma is extended toward a position angle of ${\rm PA}\sim180^{\circ}$ East of North, which notably is not along the projected directions of either the comet's anti-Solar or negative heliocentric velocity vectors.
We find a coma color of ${\rm F200W}-{\rm F277W}=-0.29\pm0.02$, which we conclude is not statistically significantly different from the ${\rm F200W}-{\rm F277W}$ coma color measured for 238P.}

\item{We find reasonable model contour fits to NIRCam images with a dynamical dust model using
$\mu$m-scale dust particles as the dominant reflectance source in F200W and F277W coma images, a nucleus obliquity of $\varepsilon\sim80^{\circ}$, and non-isotropic dust emission from the South polar region (latitudes of $<-45^{\circ}$).  In this model, we find velocities for $\mu$m-sized grains of $\sim20-40$~m~s$^{-1}$, and an average steady-state mass loss rate of $\sim0.8$~kg~s$^{-1}$ (assuming a dust grain density of $\rho_d=2500$~kg~m$^{-3}$).  This result corresponds to a $\sim2.5\times$ decline in the dust production rate between 2012-2013 and 2023-2024, consistent with photometric analysis results.
}

\item{Using observations obtained using several different ground-based facilities, we find that 358P's absolute magnitude (as measured within photometry apertures with radii of $\rho=5000$~km at the distance of the comet) brightened by about $-3.5$ mag (a factor
of $\sim25\times$) over about 150 days between UT 2023 July 26 ($\nu=328.8^{\circ}$) and UT 2023 December 21 ($\nu=11.9^{\circ}$) when the comet reached its brightest absolute magnitude, and then faded by about $0.5$~mag (about a $35$\% decline) over the next $\sim100$ days between UT 2023 December 21 and
UT 2024 March 02 ($\nu=32.6^{\circ}$). The long duration of activity based on photometry is consistent with the photometric behavior reported for 358P's 2012-2013 active apparition, although the comet's intrinsic brightness was observed to peak at a slightly later orbit position ($\nu\sim20^{\circ}$) in 2012-2013 compared to 2023-2024. Meanwhile, $Af\rho$ measurements from ground-based photometry indicate that the dust production rate has declined by a factor of $\sim2.5$ between 2012-2013 and 2023-2024, consistent with dust modeling results.}
\item{Using dust production rates derived from published dust modeling results and our own dust modeling analysis, we estimate 358P's dust-to-gas ratio to be $Q_{\rm dust}/Q_{\rm H_2O}\sim0.5$, which we note is highly uncertain due to modeling assumptions made for the values of several parameters, but reasonably consistent with the ratio of $Q_{\rm dust}/Q_{\rm H_2O}\sim0.8$ (using a grain density of $\rho_d=2500$~kg~m$^{-3}$) found for previous \jwst{} observations of 238P, given the many assumptions made.  Using $Af\rho$ as an alternative dust production rate proxy, we obtain $\log_{10}(Af\rho/Q_{\rm H_2O})=-24.8\pm0.2$ for 358P, which again is reasonably consistent with the value of $\log_{10}(Af\rho/Q_{\rm H_2O})=-24.4\pm0.2$ found for 238P, as well as for other comets from the literature at similar heliocentric distances.  Together, these results suggest that reasonable first-order estimates of the water production rates of MBCs for which only ground-based optical photometry data are available could be estimated by assuming $Q_{\rm dust}/Q_{\rm H_2O}\sim1$ or $\log_{10}(Af\rho/Q_{\rm H_2O})\sim-24.6$.}
\end{enumerate}

\section*{Data and Software Availability\label{section:software}}

Pipeline-processed \jwst{} data are publicly available from the Space Telescope Science Institute's Mikulski Archive for Space Telescopes at \url{https://mast.stsci.edu/} under JWST program ID 4250 and at \url{https://doi.org/10.17909/3w17-bx08}.  Ground-based image data from Gemini Observatory are publicly available online from the Gemini Observatory Archive at \url{https://archive.gemini.edu/} under program IDs GN-2022B-Q-307, GN-2024A-Q-112, and GS-2022B-Q-111.

This work makes use of the Planetary Spectrum Generator at \url{https://psg.gsfc.nasa.gov/} \citep{villanueva2018_psg}, the Ice Sublimation Model at \url{https://github.com/Small-Bodies-Node/ice-sublimation}, and the \jwst{} science data calibration pipeline at \url{https://github.com/spacetelescope/jwst/}.
This research also makes use of the Jet Propulsion Laboratory's Horizons online ephemeris generation tool \citep{giorgini1996_horizons};
NASA's Astrophysics Data System Bibliographic Services\footnote{\url{https://ui.adsabs.harvard.edu/}}, which is funded by NASA under Cooperative Agreement 80NSSC21M00561;
{\tt astropy}, a community-developed core {\tt python} package for astronomy \citep{astropy2018_astropy}; {\tt ccdproc}, an {\tt astropy} package for image reduction \citep{craig2023_ccdproc};
{\tt L.A.Cosmic}, a cosmic ray rejection algorithm \citep{vandokkum2001_lacosmic};
{\tt pyraf} \citep{stsci2012_pyraf_new}, a product of the Space Telescope Science Institute, which is operated by AURA for NASA; {\tt sbpy}, an {\tt astropy} affiliated package for small-body planetary astronomy \citep{mommert2019_sbpy};
a {\tt python} implementation of the \citet{cowan1979_cometsublimation} sublimation model \citep{vanselous21_ice_e20745b}; and
{\tt uncertainties} (version 3.0.2), a {\tt python} package for calculations with uncertainties by E.~O.\ Lebigot\footnote{\url{http://pythonhosted.org/uncertainties/}}.

\acknowledgments

We thank two anonymous reviewers for useful suggestions that helped to improve this manuscript.
This work benefited from support from the International Space Science Institute, Bern, Switzerland, through the hosting and provision of financial support for an international team to discuss the science of main-belt comets.  This work is based on observations made with the NASA/ESA/CSA James Webb Space Telescope.  The data were obtained from the Mikulski Archive for Space Telescopes at the Space Telescope Science Institute, which is operated by the Association of Universities for Research in Astronomy, Inc., under NASA contract NAS 5-03127 for \jwst{}.  These observations are associated with \jwst{} General Observer Program 4250.
Support for this work was provided to H.H.H., J.W.N., M.S.P.K., and D.B.\ by NASA through a grant from the Space Telescope Science Institute, which is operated by the Association of Universities for Research in Astronomy, Inc., under contract NAS 5-26555.
H.H.H., C.O.C., M.S.P.K., M.M.K., W.J.O., J.P., S.S.S., A.T., and C.A.T.\ also acknowledge support from the NASA Solar System Observations program (Grant 80NSSC19K0869).  The work of J.P.\ was conducted at the Jet Propulsion Laboratory, California Institute of Technology, under a contract with the National Aeronautics and Space Administration (80NM0018D0004).

The authors thank C.\ Soto, B.\ Hilbert, K.\ Glidic and other Space Telescope Science Institute staff for their assistance in obtaining \jwst{} observations,
and acknowledge B.~N.~L.\ Sharkey for useful discussion regarding the analysis of NIRSpec reflectance spectra, and N.\ Sch{\"o}rghofer and Y.\ Zhang for useful discussions regarding thermal modeling analyses of MBCs.

This work is based in part on observations obtained at the international Gemini Observatory (under Programs GN-2022B-Q-307, GN-2024A-Q-112, and GS-2022B-Q-111), a program of NSF NOIRLab, which is managed by the Association of Universities for Research in Astronomy (AURA) under a cooperative agreement with the U.S. National Science Foundation on behalf of the Gemini Observatory partnership: the U.S. National Science Foundation (United States), National Research Council (Canada), Agencia Nacional de Investigaci\'{o}n y Desarrollo (Chile), Ministerio de Ciencia, Tecnolog\'{i}a e Innovaci\'{o}n (Argentina), Minist\'{e}rio da Ci\^{e}ncia, Tecnologia, Inova\c{c}\~{o}es e Comunica\c{c}\~{o}es (Brazil), and Korea Astronomy and Space Science Institute (Republic of Korea).
Portions of this work were specifically enabled by observations made from the Gemini North telescope, located within the Maunakea Science Reserve and adjacent to the summit of Maunakea. We are grateful for the privilege of observing the Universe from a place that is unique in both its astronomical quality and its cultural significance.

This work is also based in part on observations obtained at the Hale Telescope at Palomar Observatory as part of a continuing collaboration between the California Institute of Technology, NASA/JPL, Yale University, and the National Astronomical Observatories of China.

This work is also based in part on observations obtained at the Southern Astrophysical Research (SOAR) telescope under program (2023A-396684), which is a joint project of the Minist\'{e}rio da Ci\^{e}ncia, Tecnologia e Inova\c{c}\~{o}es (MCTI/LNA) do Brasil, the US National Science Foundation's NOIRLab, the University of North Carolina at Chapel Hill (UNC), and Michigan State University (MSU).

This work is also based in part on observations obtained at the Lowell Discovery Telescope at Lowell Observatory.
Lowell is a private, non-profit institution dedicated to astrophysical research and public appreciation of astronomy and operates the LDT in partnership with Boston University, the University of Maryland, the University of Toledo, Northern Arizona University and Yale University.
The Large Monolithic Imager was built by Lowell Observatory using funds provided by the National Science Foundation (AST-1005313). The University of Maryland LDT observing team consists of Q.\ Ye, J.\ Bauer, A.\ Gicquel-Brodtke, T.\ Farnham, L.\ Farnham, C.\ Holt, M.\ S.\ P.\ Kelley, J.\ Kloos, and J.\ Sunshine.
They thank C.\ Siqueiros and other LDT staff for assistance in operating the telescope.

The authors thank 
T.\ Barlow, P.\ Nied, D.\ Roderick, J.\ Stone, I.\ Wilson, and other Palomar Observatory staff; 
J.\ Ball, J.\ Berghuis, R.\ Carrasco, K.\ Chiboucas, B. Cooper, J.\ Ferrara, C.\ Figura, G.\ Gimeno, J.-E.\ Heo, D.\ Hung, Y.\ Kang, S.\ Leggett, P.\ Mart\'in-Ravelo, C.\ Mart\'inez-V\'azquez, J.\ Miller, T.\ Mocnik, M. Rawlings, K.\ Silva, A.\ Smith, S.\ Stewart, S.\ Xu, and other Gemini Observatory staff; and
E.\ Bull, A.\ Grove, M.\ Hirose, J.\ Luisi, H.\ Sinclair-Wentworth, K.\ Southon, F.\ Gunn, and other Mount John Observatory staff for their assistance in obtaining observations.

The authors thank S.\ Points, C.\ Corco, R.\ Aviles, and other SOAR staff for their assistance in obtaining observations.






\clearpage

\bibliography{main.bbl}{}
\bibliographystyle{aasjournal}

\clearpage



\renewcommand{\thesubsection}{\Alph{subsection}}

\appendix
\setcounter{figure}{0}
\renewcommand{\thefigure}{A\arabic{figure}}

\section{Composite Images from Ground-based Observations\label{section:appendix_groundbased_images}}

\begin{figure*}[ht]
    \centering
    \includegraphics[width=0.8\linewidth]{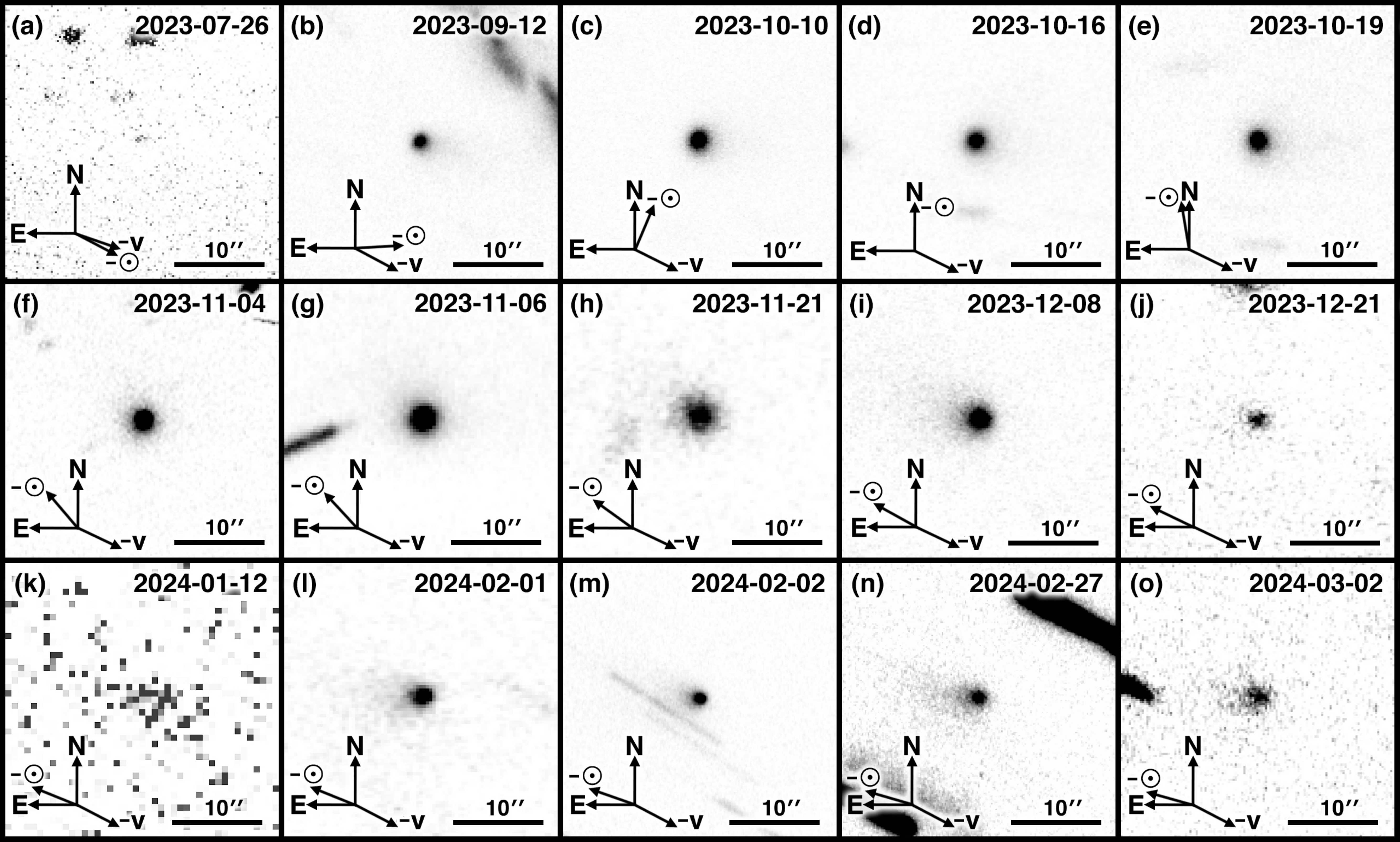}
    \caption{Composite images constructed from ground-based optical observations of 358P listed in Table~\ref{table:ground_observations_358p}.}
    \label{fig:358p_optical_images}
\end{figure*}

\begin{figure*}[ht]
    \centering
    \includegraphics[width=0.64\linewidth]{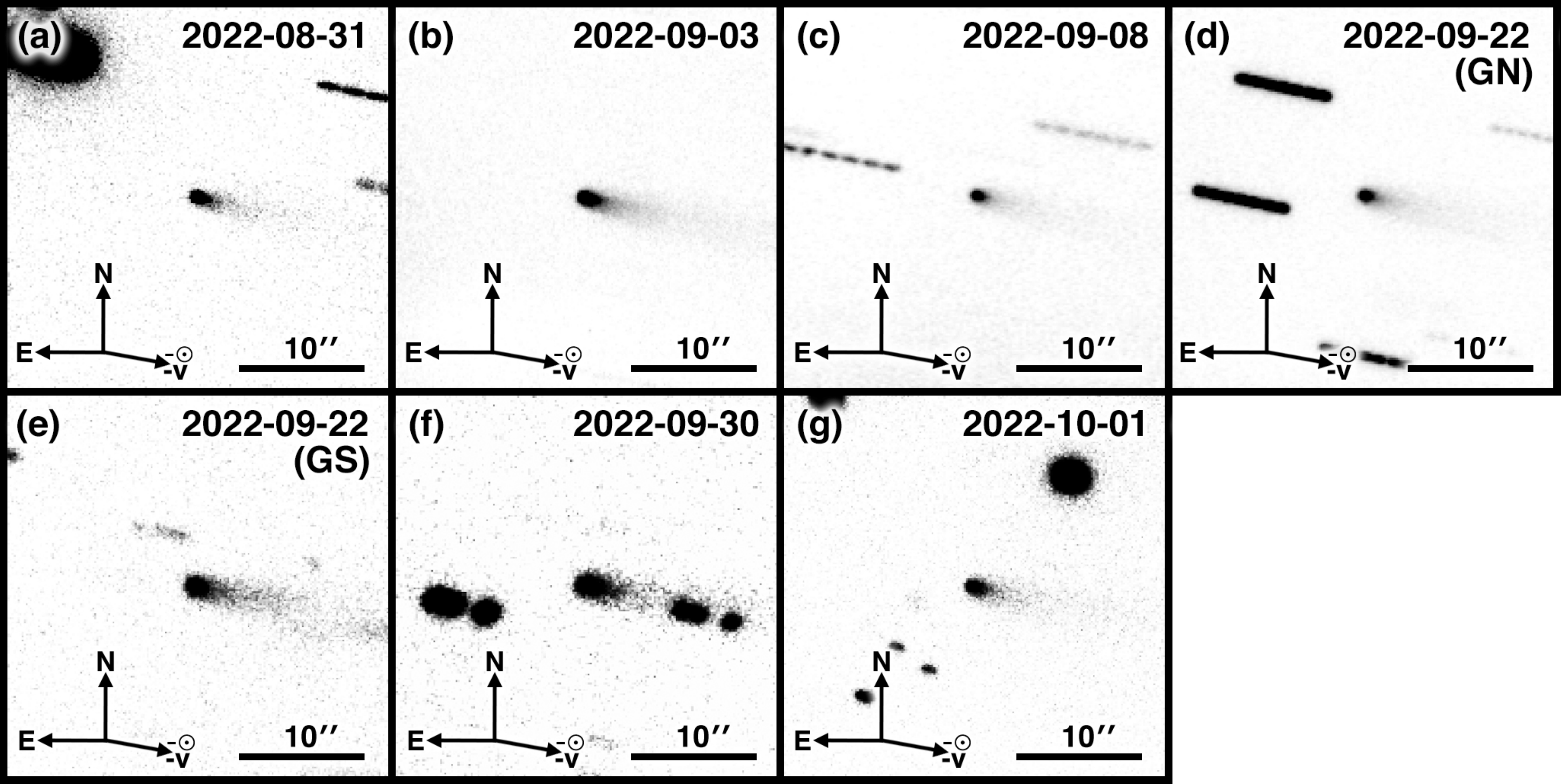}
    \caption{Composite images constructed from ground-based optical observations of 238P listed in Table~\ref{table:ground_observations_238p}.}
    \label{fig:238p_optical_images}
\end{figure*}

\clearpage

\section{Reported Activity Observations of Recurrently Active MBCs}\label{section:appendix_recurrent_activity}

In this Appendix, we show the true anomaly and heliocentric distance ranges of visible activity (i.e., dust comae, tails, and/or trails) from reported observations for all confirmed recurrently active MBCs as of 2024 October 28 (Table~\ref{table:recurrent_mbcs}; Figure~\ref{fig:active_ranges}).  We note that the absence of reported activity at particular times does not necessarily exclude the existence of activity at those times, as observations may not have occurred at those times, may not have been sensitive enough to detect faint activity at the time, or simply may not have been published yet as of the time of this writing (2024 October 28).  We also note that observations of visible dust listed here, particularly at larger true anomalies and heliocentric distances while MBCs are outbound, may not necessarily correspond to times of active dust production, as they also cover the periods of time after active dust production has ceased but before past ejected dust has fully dissipated.

\setcounter{table}{0}
\setcounter{figure}{0}
\renewcommand{\thefigure}{B\arabic{figure}}
\renewcommand{\thetable}{B\arabic{table}}

\setlength{\tabcolsep}{1pt}
\setlength{\extrarowheight}{0em}
\begin{table*}[htb!]
\caption{Reported Activity Observations of Recurrently Active MBCs}
\centering
\smallskip
\footnotesize
\begin{tabular}{lcccrccccccclclcrcrcccccc}
\hline\hline
\multicolumn{1}{c}{Object}
 &~~& \multicolumn{1}{c}{$a$$^a$}
 &~~& \multicolumn{1}{c}{$e$$^b$}
 &~~& \multicolumn{1}{c}{$i$$^c$}
 &~~& \multicolumn{1}{c}{$H_V$$^d$}
 &~~& \multicolumn{1}{c}{$r_n$$^e$}
 &~~& \multicolumn{3}{c}{Dates of reported activity$^f$}
 &~~& \multicolumn{3}{c}{$\nu$ range$^g$}
 &~~& \multicolumn{3}{c}{$r_h$ range$^h$}
 &~~& \multicolumn{1}{c}{Ref.$^i$}
 \\
\hline
133P/Elst-Pizarro$^j$            && 3.164 && 0.156 && 1.390 && 15.9 && 2.0 && 1996 Jul 14 &-& 1996 Sep 18 && 22 &:& 37 && 2.65 &:& 2.71 && [1] \\
\multicolumn{1}{c}{...} && \multicolumn{1}{c}{...} && \multicolumn{1}{c}{...} && \multicolumn{1}{c}{...} && \multicolumn{1}{c}{...} && \multicolumn{1}{c}{...} && 2002 Jul 13 &-& 2002 Dec 28 && 63 &:& 89 && 2.86 &:& 3.06 && [1]  \\
\multicolumn{1}{c}{...} && \multicolumn{1}{c}{...} && \multicolumn{1}{c}{...} && \multicolumn{1}{c}{...} && \multicolumn{1}{c}{...} && \multicolumn{1}{c}{...} && 2007 May 19 &-& 2008 Nov 27$^k$ && $-$10 &:& 109 && 2.64 &:& 3.25 && [1]  \\
\multicolumn{1}{c}{...} && \multicolumn{1}{c}{...} && \multicolumn{1}{c}{...} && \multicolumn{1}{c}{...} && \multicolumn{1}{c}{...} && \multicolumn{1}{c}{...} && 2013 Jun 4 &-& 2013 Jul 10 && 28 &:& 37 && 2.69 &:& 2.72 && [1]  \\
238P/Read$^j$                     && 3.166 && 0.252 && 1.264 && 20.5 && 0.24 && 2005 Oct 24 &-& 2007 Jan 27 && 26 &:& 123 && 2.42 &:& 3.43 && [2] \\
\multicolumn{1}{c}{...} && \multicolumn{1}{c}{...} && \multicolumn{1}{c}{...} && \multicolumn{1}{c}{...} && \multicolumn{1}{c}{...} && \multicolumn{1}{c}{...} && 2011 Aug 29 &-& 2011 Dec 31 && 50 &:& 79 && 2.54 &:& 2.82 && [2]  \\
\multicolumn{1}{c}{...} && \multicolumn{1}{c}{...} && \multicolumn{1}{c}{...} && \multicolumn{1}{c}{...} && \multicolumn{1}{c}{...} && \multicolumn{1}{c}{...} && 2016 Jul 8 &-& 2017 Jan 26 && $-$31 &:& 29 && 2.37 &:& 2.44 && [2]  \\
\multicolumn{1}{c}{...} && \multicolumn{1}{c}{...} && \multicolumn{1}{c}{...} && \multicolumn{1}{c}{...} && \multicolumn{1}{c}{...} && \multicolumn{1}{c}{...} && 2022 Aug 31 &-& 2022 Oct 1 && 26 &:& 35 && 2.42 &:& 2.46 && [2]  \\
259P/Garradd                  && 2.727 && 0.342 && 15.899 && 19.9 && 0.3 && 2008 Sep 24 &-& 2008 Nov 11 && 28 &:& 49 && 1.85 &:& 1.96 && [3] \\
\multicolumn{1}{c}{...} && \multicolumn{1}{c}{...} && \multicolumn{1}{c}{...} && \multicolumn{1}{c}{...} && \multicolumn{1}{c}{...} && \multicolumn{1}{c}{...} && 2017 Mar 28 &-& 2017 Dec 22 && $-$56 &:& 60 && 1.81 &:& 2.07 && [3] \\
288P/(300163) 2006 VW$_{139}$ && 3.050 && 0.200 && 3.238 && 17.1 && 0.9, 0.6 && 2000 Sep 3 &-& 2000 Nov 17 && $-$21 &:& 0 && 2.46 &:& 2.49 && [4]  \\
\multicolumn{1}{c}{...} && \multicolumn{1}{c}{...} && \multicolumn{1}{c}{...} && \multicolumn{1}{c}{...} && \multicolumn{1}{c}{...} && \multicolumn{1}{c}{...} && 2011 Aug 30 &-& 2012 Feb 21 && 12 &:& 59 && 2.45 &:& 2.65 && [4]  \\
\multicolumn{1}{c}{...} && \multicolumn{1}{c}{...} && \multicolumn{1}{c}{...} && \multicolumn{1}{c}{...} && \multicolumn{1}{c}{...} && \multicolumn{1}{c}{...} && 2016 Jun 8 &-& 2017 Jan 15 && $-$42 &:& 19 && 2.44 &:& 2.55  && [4] \\
313P/Gibbs                    && 3.155 && 0.242 && 10.966 && 17.8 && 0.9 && 2003 Sep 30 &-& 2003 Nov 28 && 31 &:& 47 && 2.43 &:& 2.53 && [5] \\
\multicolumn{1}{c}{...} && \multicolumn{1}{c}{...} && \multicolumn{1}{c}{...} && \multicolumn{1}{c}{...} && \multicolumn{1}{c}{...} && \multicolumn{1}{c}{...} && 2014 Aug 6 &-& 2015 Mar 5 && $-$6 &:& 53 && 2.39 &:& 2.59 && [5] \\
324P/La Sagra                 && 3.094 && 0.154 && 21.420 && 18.6 && 0.6 && 2010 Aug 16 &-& 2011 Aug 31 && 13 &:& 96 && 2.63 &:& 3.07 && [6] \\
\multicolumn{1}{c}{...} && \multicolumn{1}{c}{...} && \multicolumn{1}{c}{...} && \multicolumn{1}{c}{...} && \multicolumn{1}{c}{...} && \multicolumn{1}{c}{...} && 2015 Mar 21 &-& 2016 Dec 29 && $-$60 &:& 89 && 2.71 &:& 3.01 && [6] \\
\multicolumn{1}{c}{...} && \multicolumn{1}{c}{...} && \multicolumn{1}{c}{...} && \multicolumn{1}{c}{...} && \multicolumn{1}{c}{...} && \multicolumn{1}{c}{...} && 2021 May 10 &-& 2021 Nov 29 && 1 &:& 50 && 2.62 &:& 2.75 && [6] \\
358P/PANSTARRS$^j$                && 3.147 && 0.239 && 11.060 && 20.2 && 0.3 &&  2012 Oct 6 &-& 2013 Feb 4 && 7 &:& 41 && 2.41 &:& 2.53  && [7] \\
\multicolumn{1}{c}{...} && \multicolumn{1}{c}{...} && \multicolumn{1}{c}{...} && \multicolumn{1}{c}{...} && \multicolumn{1}{c}{...} && \multicolumn{1}{c}{...} && 2017 Nov 18 &-& 2017 Dec 18 && $-$41 &:& $-$33 && 2.48 &:& 2.52 && [7] \\
\multicolumn{1}{c}{...} && \multicolumn{1}{c}{...} && \multicolumn{1}{c}{...} && \multicolumn{1}{c}{...} && \multicolumn{1}{c}{...} && \multicolumn{1}{c}{...} && 2023 Jul 26 &-& 2024 Mar 2 && $-$31 &:& 33 && 2.40 &:& 2.47 && [7]  \\
432P/PANSTARRS                && 3.037 && 0.243 && 10.070 && ? && ? && \multicolumn{3}{c}{2016 Oct 25} && \multicolumn{3}{c}{48} && \multicolumn{3}{c}{2.46} && [8] \\
\multicolumn{1}{c}{...} && \multicolumn{1}{c}{...} && \multicolumn{1}{c}{...} && \multicolumn{1}{c}{...} && \multicolumn{1}{c}{...} && \multicolumn{1}{c}{...} && 2021 Aug 9 &-& 2021 Aug 10 && $-$7 &:& $-$6 && \multicolumn{3}{c}{2.30} && [8] \\
433P/(248370) 2005 QN$_{173}$ && 3.066 && 0.225 &&  0.068 && 16.4 && 1.6 && \multicolumn{3}{c}{2016 Jul 22} && \multicolumn{3}{c}{57} && \multicolumn{3}{c}{2.59} && [9] \\
\multicolumn{1}{c}{...} && \multicolumn{1}{c}{...} && \multicolumn{1}{c}{...} && \multicolumn{1}{c}{...} && \multicolumn{1}{c}{...} && \multicolumn{1}{c}{...} && 2021 Jul 7 &-& 2021 Dec 26 && 16 &:& 63 && 2.39 &:& 2.64 && [9] \\
456P/PANSTARRS$^l$     && 3.165 && 0.119 && 16.962 && 17.5 && 1.0 && 2021 Jun 9 &-& 2021 Jun 10 && 124 &:& 125 && \multicolumn{3}{c}{3.35} && [10] \\
\multicolumn{1}{c}{...} && \multicolumn{1}{c}{...} && \multicolumn{1}{c}{...} && \multicolumn{1}{c}{...} && \multicolumn{1}{c}{...} && \multicolumn{1}{c}{...} && 2024 Oct 3 &-& 2024 Oct 26 && $-$42 &:& $-$37 && 2.86 &:& 2.88 && [10]  \\
457P/Lemmon-PANSTARRS$^{j,m}$         && 2.646 && 0.120 &&  5.224 && 19.3 && 0.4 && \multicolumn{3}{c}{2016 Jul 3} && \multicolumn{3}{c}{48} && \multicolumn{3}{c}{2.42} && [11] \\
\multicolumn{1}{c}{...} && \multicolumn{1}{c}{...} && \multicolumn{1}{c}{...} && \multicolumn{1}{c}{...} && \multicolumn{1}{c}{...} && \multicolumn{1}{c}{...} && 2020 May 3 &-& 2020 Nov 25 && 0 &:& 58 && 2.33 &:& 2.45 && [11] \\
477P/PANSTARRS                && 3.007 && 0.416 &&  8.909 && $>$18.2 && $<$0.7 && 2018 Aug 8 &-& 2018 Dec 28 && $-$31 &:& 38 && 1.76 &:& 1.88 && [12] \\
\multicolumn{1}{c}{...} && \multicolumn{1}{c}{...} && \multicolumn{1}{c}{...} && \multicolumn{1}{c}{...} && \multicolumn{1}{c}{...} && \multicolumn{1}{c}{...} && 2023 Nov 15 &-& 2024 Jan 8 && $-$21 &:& 6 && 1.75 &:-& 1.79 && [12] \\
483P/PANSTARRS-A              && 3.172 && 0.228 && 14.330 && $>$20.0 && $<$0.3 && 2010 Apr 12 &-& 2010 Apr 17 && $-$56 &:& $-$54 && 2.63 &:& 2.64 && [13] \\
\multicolumn{1}{c}{...} && \multicolumn{1}{c}{...} && \multicolumn{1}{c}{...} && \multicolumn{1}{c}{...} && \multicolumn{1}{c}{...} && \multicolumn{1}{c}{...} && 2016 May 5 &-& 2016 Aug 4 && $-$14 &:& 12 && 2.45 &:& 2.46 && [13] \\
\multicolumn{1}{c}{...} && \multicolumn{1}{c}{...} && \multicolumn{1}{c}{...} && \multicolumn{1}{c}{...} && \multicolumn{1}{c}{...} && \multicolumn{1}{c}{...} && 2022 Apr 8 &-& 2022 Aug 2 && 13 &:& 45 && 2.46 &:& 2.59 && [13] \\
483P/PANSTARRS-B              && 3.172 && 0.228 && 14.331 && $>$21.3 && $<$0.2 && 2010 Apr 12 &-& 2010 Apr 17 && $-$56 &:& $-$54 && 2.63 &:& 2.64 && [13] \\
\multicolumn{1}{c}{...} && \multicolumn{1}{c}{...} && \multicolumn{1}{c}{...} && \multicolumn{1}{c}{...} && \multicolumn{1}{c}{...} && \multicolumn{1}{c}{...} && 2016 May 5 &-& 2016 Aug 4 && $-$14 &:& 12 && 2.45 &:& 2.46 && [13] \\
\multicolumn{1}{c}{...} && \multicolumn{1}{c}{...} && \multicolumn{1}{c}{...} && \multicolumn{1}{c}{...} && \multicolumn{1}{c}{...} && \multicolumn{1}{c}{...} && 2021 May 29 &-& 2021 May 31 && $-$70 &:& $-$69 && 2.78 &:& 2.79 && [13] \\
2010 LH$_{15}$$^n$            && 2.744 && 0.355 && 10.906 && 18.8 && 0.5 && 2010 Sep 27 &-& 2010 Oct 10 && 22 &:& 28 && 1.79 &:& 1.81 && [14] \\
\multicolumn{1}{c}{...} && \multicolumn{1}{c}{...} && \multicolumn{1}{c}{...} && \multicolumn{1}{c}{...} && \multicolumn{1}{c}{...} && \multicolumn{1}{c}{...} && 2019 Aug 10 &-& 2019 Nov 3 && $-$14 &:&  27 && 1.77 &:& 1.82 && [14] \\
\hline
\hline
\multicolumn{25}{l}{$^a$Semimajor axis, in au.} \\
\multicolumn{25}{l}{$^b$Eccentricity.} \\
\multicolumn{25}{l}{$^c$Inclination, in degrees.} \\
\multicolumn{25}{l}{$^d$$V$-band absolute magnitude, from \citet{hsieh2023_mbcnuclei}, unless otherwise specified.} \\
\multicolumn{25}{l}{$^e$Estimated effective nucleus radius (assuming $p_V=0.05$), in km, from \citet{hsieh2023_mbcnuclei}, unless otherwise specified.} \\
\multicolumn{25}{l}{$^f$Date range of activity reported as of 2024 Oct 21.} \\
\multicolumn{25}{l}{$^g$True anomaly range of reported activity, in degrees.} \\
\multicolumn{25}{l}{$^h$Heliocentric distance range of reported activity, in au.} \\
\multicolumn{25}{l}{$^i$[1]
 \citet{elst1996_133p}, \citet{hammergren1996_133p},
 \citet{hsieh2004_133p,hsieh2010_133p,hsieh2013_133p}, \citet{lowry2005_whtcomets}, \citet{kaluna2011_133pactivity},} \\
\multicolumn{25}{l}{~\citet{jewitt2014_133p}; [2] \citet{read2005_238p}, \citet{hsieh2009_238p,hsieh2018_238p288p}, this work; 
  [3] \citet{garradd2008_259p}, \citet{jewitt2009_259p},} \\
\multicolumn{25}{l}{~\citet{hsieh2021_259p}; [4] \citet{hsieh2012_288p,hsieh2018_238p288p};
  [5] \citet{hsieh2015_313p}, \citet{jewitt2015_313p2};
  [6] \citet{hsieh2012_324p},} \\
\multicolumn{25}{l}{~\citet{hsieh2015_324p}, \citet{mastropietro2024_324p}; [7] \citet{hsieh2013_p2012t1,hsieh2018_358p}, this work;
  [8] \citet{ramanjooloo2021_432p},} \\
\multicolumn{25}{l}{~\citet{weryk2021_432p}; [9] \citet{hsieh2021_433p}, \citet{chandler2021_433p}, \citet{ivanova2023_433p}; [10] \citet{wainscoat2021_p2021l4},} \\
\multicolumn{25}{l}{~\citet{hsieh2024_456p} [11] \citet{weryk2020_p2020o1},\citet{kim2022_p2020o1}, \citet{ly2023_457p_cbet}; [12] \citet{weryk2018_p2018p3}, \citet{kim2022_p2018p3},} \\
\multicolumn{25}{l}{~\citet{oribe2024_p2018p3_cbet}, \citet{abarzuza2024_477p_atel};[13] \citet{weryk2016_p2016j1_cbet}, \citet{moreno2017_p2016j1}, \citet{hui2017_p2016j1},} \\
\multicolumn{25}{l}{~\citet{hsieh2023_mbcnuclei,hsieh2024_p2016j1_cbet}; [14] \citet{chandler2023_2010LH15},\citet{mastropietro2024_2010lh15}.} \\
\multicolumn{25}{l}{$^j$Observed by \jwst{}.} \\
\multicolumn{25}{l}{$^k$Estimated date; exact date not published in \citet{kaluna2011_133pactivity}.} \\
\multicolumn{25}{l}{$^l$Nucleus size from \citet{hsieh2024_456p}.} \\
\multicolumn{25}{l}{$^m$Nucleus size from \citet{kim2022_p2020o1}.} \\
\multicolumn{25}{l}{$^n$Nucleus size from \citet{mastropietro2024_2010lh15}.} \\
\end{tabular}
\label{table:recurrent_mbcs}
\end{table*}

\begin{figure*}[ht]
    \centering
    \includegraphics[width=0.97\linewidth]{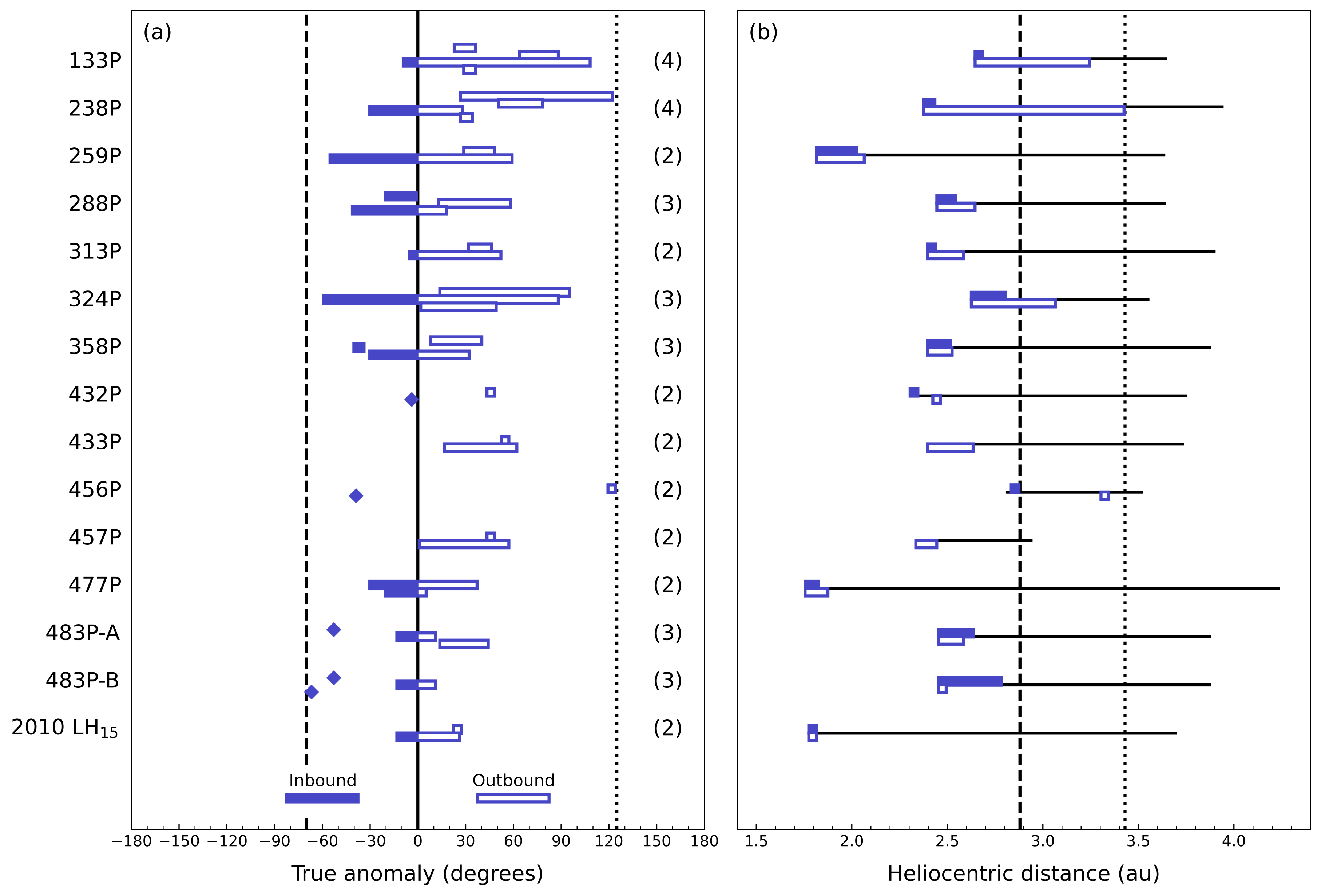}
    \caption{Active ranges, extending between the earliest and latest observations for which activity has been reported during individual apparitions (Table~\ref{table:recurrent_mbcs}), in terms of (a) true anomaly and (b) heliocentric distance for confirmed recurrently active MBCs. Solid blue line segments indicate the inbound (pre-perihelion) portion of each object's orbit while outlined blue line segments indicate the outbound (post-perihelion) portion of each object's orbit. 
    In (a), individual active apparitions are plotted separately with apparitions for the same objects grouped together, with the earliest apparitions at the top and latest apparitions at the bottom of each grouping. Perihelion is marked with a solid vertical line, while the earliest and latest orbit positions in terms of true anomaly at which activity has been observed for any MBC (in this case, for 483P-B and 456P, respectively) are marked with a dashed and dotted vertical line, respectively. 
    Numbers in parentheses on the right side indicate the number of active apparitions for which observations have been reported for each object as of 2024 Oct 28.  
    In (b), solid and outlined blue line segments represent the combined ranges of reported observations from all active apparitions of each object,
    with horizontal black line segments indicating the heliocentric distance range covered by the orbit of each object. The most distant positions at which activity has been observed for any MBC inbound to and outbound from perihelion (in this case, for 324P and 238P, respectively) are marked with a dashed and dotted vertical line, respectively. After \citet{hsieh2018_238p288p}.}
    \label{fig:active_ranges}
\end{figure*}



\end{CJK*}
\end{document}